\documentclass[11pt]{article}
\usepackage{amssymb}

\usepackage{a4}

\usepackage{graphicx}

\usepackage{endnotes}

\usepackage[stable]{footmisc}
\newcommand{\dd}{\mathrm{d} }
\newcommand{\St}{\mathrm{I}}

\begin{document}

\author{\Large John Hayward \normalsize \\\\
School of  Computing and Mathematics\\ University of South Wales\\
Pontypridd, CF37 1DL, Wales, UK \\\\
john.hayward@southwales.ac.uk\footnote{Alternative contact:  www.churchmodel.org.uk,  jhayward@churchmodel.org.uk}}
\title{Mathematical Modeling of Church Growth: \\   A System Dynamics Approach}

\date{2018 (1999)}
\maketitle

\begin{abstract}
The possibility of using mathematics to model church growth is
investigated using ideas from population modeling. It is proposed
that a major mechanism of growth is through contact between
religious enthusiasts and unbelievers, where the enthusiasts are
only enthusiastic for a limited period. After that period they
remain church members but less effective in recruitment. This
leads to the general epidemic model which is applied to a variety
of church growth situations. Results show that even a simple model
like this can help understand the way in which churches grow,
particularly in times of religious revival. This is a revised version of Hayward (1999) using System Dynamics and some small modifications to the SIR model \endnote{This paper is a revised version of  Hayward (1999), which was the first published church growth model. The main revision is that the model is expressed in the system dynamics diagrammatic notation of  Forrester (1961). This style of modeling  gives a clearer connection between  assumptions and the causal structure of a model; and exposes the feedback in a model, relating it to model behavior. Thus this revised paper differs from Hayward (1999) in the model construction. The paper uses recent developments in measuring the effects of feedback loops (Hayward \& Boswell, 2014), and the identification of causal effects of variables on each other as social forces. This material is added to section 5: \emph{Model Results}   (originally named \emph{Numerical Solutions} in 1999).

\hspace{10pt}The original 1999 paper used the Crowd Model of the spread of disease, sometimes called the Mass Action model. Research since 1999 has shown the Fixed Contacts model is more appropriate for word-of-mouth diffusion as density effects are less pronounced than for a physical disease. In this revised paper the Fixed Contacts model is used throughout. The Fixed Contacts model is sometimes called Standard Incidence.

\hspace{10pt}The year after the 1999 paper was published, Hayward (2000) added a small modification to the model so that not all new converts become enthusiasts, those who spread the religion. It is this model that has come to be known as the Limited Enthusiasm model, thus this modification is used in this revised paper.}.

\end{abstract}




\section{Introduction}

\subsection{Background To Church Growth}

\label{Background}

``Church Growth'' is a recent subject area which seeks to analyse why
Christian churches, at various levels of organisation, grow or decline.
Although spiritual growth is included in the subject, numerical growth -- how
many attend or belong to a church -- is a vital area for analysis. The
stimulus for investigating the reasons why churches grow came from western
missionary organisations in the late 1950's. They were concerned with the
effectiveness of the missions they had founded, and needed to examine this
effectiveness in order to determine priorities for funding. The pioneer in
this field was Donald McGavran who did much to encourage these missionary
organisations to see the vast potential for numerical growth in the
non-western world. Although numerical growth is not a requirement of all
missionary endeavours (McGavran, 1963), it was nevertheless felt that
situations needed to be analysed with the factors that encourage and
inhibit growth identified.

Church growth thinking is divided into two strands: the Church Growth
Movement, which is based within the denominations and seminaries of the
Christian church to serve their needs; and the Social Science strand whose
focus is academic research. The two strands have tended to remain separate,
perhaps reflecting a certain amount of mutual suspicion between them. It is
perhaps not surprising that those working within churches distrust
sociologists as until recently the prevailing social science view was that
religion had no significant place in modern society and would die out -- a
view often called secularisation theory (Stark and Bainbridge, 1987, pp.13--14; Warner, 1993)  The comments of the anthropologist Wallace (1966, p.265) are typical:
\emph{``the evolutionary future of religion is extinction''}, as are Berger's remarks that assertions of
supernaturalism would be restricted to smaller groups or backward regions
(Berger, 1970). From the sociological point of view the church growth
strand could also be regarded as suspect, since it is not neutral and often
lacks academic rigour.

Many of the principles of the Church Growth Movement were developed at The
Fuller Theological Seminary in Pasadena, California, where Donald McGavran
became professor of Church Growth. This movement has itself grown over the
years with numerous organisations teaching church growth principles, acting
as consultants to denominations and local domestic churches as well as
mission organisations. In the USA there is the North American Church Growth
Association, with a membership from across the denominations, which
encourages this way of thinking through its literature. Similar work is
undertaken in other countries. Much of the work is qualitative, with
quantitative work restricted to data gathering, interpretation and
application to churches in the way that business consultants might advise
firms. Brierley (1991) in the UK is typical. There is
however little attempt at general theories of church growth, only heuristic
principles.

The Social Science strand grew very much as a reaction to a key,
but controversial, book by Dean Kelley, originally published 1972 and revised in 1986,   which put forth an explanation as to why
conservative churches are strong. Kelley's thesis, stated simply,
was that conservatives churches are strong and hence grow, whereas
the more liberal churches decline. This has led to a flourish of
research to either prove or disprove this thesis (Hoge and Roozen, 1979; Roozen and Hadaway, 1993).

This history of the two strands, and their relationship, is described by Inskeep (1993). One clear distinction between them has emerged: The
church growth work tends to view growth mainly influenced by factors within the
churches themselves -- institutional factors; whereas the social science
strand views growth as primarily determined by conditions in the surrounding
society -- contextual factors. However the common factor of both strands is
their desire to understand how churches grow.

Numerous authors have noted that in the USA the Christian churches, as well
as other religions, continue to grow despite the predictions of
secularisation theory. This has led to the beginnings of a paradigm shift in
thinking from secularisation theory, as typified by Berger (1969), towards one which sees religions flourishing in what is
essentially an open market religious economy. This fundamental
change is described by Warner (1993), and is
typified by the work of Stark (Stark, 1996; Stark and Bainbridge, 1985; 1987; Fink and Stark 1992) and Iannaccone (Iannaccone, 1992; 1994;  Iannaccone, Olson, and Stark, 1995),    among others. Indeed Iannaccone, using a model
based on rational choice theory, affirms Kelley's thesis that
strictness, makes churches strong, even in modern society. This
has implications for the study of church growth as it becomes
increasingly accepted that religious revivals are not only facts
of history, but continue to take place in modern society among all
classes (Stark and Bainbridge, 1985, ch.9; Stark and Iannaccone, 1994; Warner, 1993, pp.1046--1048).

Much of church growth modelling tends to be statistical, Doyle and Kelley (1979) is typical. This approach is essentially empirical in nature.
The question can be asked if any theoretical understanding could be brought
into the situation that might help explain why the figures behave as they
do. Theories have been expressed qualitatively and tested against data (for
example see Hoge (1979). Stark comes closer to a theory by
computing arithmetically the implications of exponential growth 
(Stark and Bainbridge, 1985, ch.16; Stark, 1996, p.7). More recently Iannaccone et. al. (1995) have produced a theory of church growth based on the
variables of time and money using economic production functions. However
none of these approaches attempts to model the dynamics of church growth in
terms of the underlying causes. The main aim of this paper is to produce
such a model of church growth, using mathematics, which will describe the
dynamics of the growth process. It is hoped that such models will give a
deeper understanding of the way in which churches grow.

\subsection{Types of Models}

\label{Types of Models}

The next step is to consider what sort of models should be developed.
Stochastic models are closer to the truth, but more difficult to handle. For
that reason deterministic models are best investigated first. Models can be
developed to investigate age profiles of the church as well as its
geographical spread, however these are unnecessary complications for an
initial model. Instead, in this paper, the only feature modeled will be that
of the change of numbers in the church over a period of time.

Rather than develop a new model from scratch, it is worth investigating if
there are similar behaviour patterns to church growth in other areas of
population modeling. This paper looks at the application of epidemic models
to church growth. These models prove useful because of the similarities
between the spread of a disease and the spread of beliefs which ultimately
leads to growth in the church. These similarities may be summarised:

\begin{itemize}
\item  There are at least two categories of people: those who have the
disease -- or belief, in the church growth case -- and those who do not.

\item  Beliefs, like many diseases, are often spread by some sort of contact
between the two categories of people. In the case of diseases the contact
may be physical, or via some intermediary mechanism such as airborne
droplets. For beliefs the contact is via oral communication.

\item  The church has frequently experienced the type of rapid growth
followed by slower periods of change typical of epidemics. In the churches
this is usually referred to as religious ``revival''. For example in Wales,
UK in 1904--5 100,000 people were added to the main Welsh denominations
(Evans, 1969, p.146), only to be followed by a period of slower growth
and eventual decline. Over a longer period South and Latin America, Africa
and some Asian countries have seen a huge growth in the churches this
century, which shows no sign of slowing down.

\item  During times of revival people are noticeably different, particularly
in regard to their enthusiasm to communicate their beliefs to others.
Their behaviour is affected. This has no doubt been a factor in the rapid
growth of the church during revivals. In Wales in 1904 such people were said
to ``have the revival''  (Lloyd-Jones 1984. pp.60--61) as if it were a
disease that could be caught! Those involved in revivals have described them
as ``contagious'', being spread from congregation to congregation 
 (Edwards, 1990, p.89).
\end{itemize}

Thus models of the spread of infectious diseases -- epidemic models -- should
prove a useful starting point to model the change in numbers in the church
over a period of time.

\subsection{Diffusion in Populations}

The diffusion process can occur in wide variety of physical, biological and
social systems. As such there is a wealth of literature covering models in
these areas which may be deterministic or stochastic and may include spatial
spread. The model in this paper is in the style of deterministic non-spatial
modelling. Banks (1994)  and Murray (1989) 
review a range of such mathematical models and their applications. In the
case of church growth the religious belief is being diffused through a
population. Thus church growth is a form of social diffusion. Early
mathematical models of social diffusion were studied by Coleman (1964) and applied to the spread of medical innovations. Kumar and Kumar (1992) and Mahajan et. al. (1990) review
more recent work. Sociological models of innovation diffusion are described
non-mathematically by Rogers (1995).

Most of the above models are variations on the logistic model of population
growth. These models assume that those possessing the innovation (adopters)
are responsible for its spread through contact with those without (potential
adopters). However the models also assume that adopters continue to spread
the innovation until it is adopted by all potential adopters, although the
coefficient of influence may decline. This will be deemed to restrictive for
modelling the spread of religion, as enthusiasm for spreading the faith not
only wanes but effectively ceases to exist for many within the church. The
fact that religious belief never spreads throughout a population lends
weight to the need to limit the process of spread. Thus church growth
modelling will be social diffusion where the enthusiasm to spread the
``innovation'' by those in possession of it is limited in duration. This
leads to a third category of people who are removed from the spreading
process. The need for this dropping-out effect in social diffusion was noted
by Webber (1972, p.231), and by Granoveter and Soong (1983) in the context of the spread of fashion, rumours and
riots. In Granoveter's model the drop-out was determined by a threshold,
with the adopters giving up the adoption. In the church growth case the
drop-out will be determined by a period of time, following Webber (1972), with the adopters remaining in the church but now
ineffective in spreading the innovation. This is the epidemic model.

The use of the epidemic model in social diffusion was proposed by
Bartholomew (1967, ch.8) to model stochastically the
spread of a rumour through a population. The model was extended by Sharif
and Ramanathan (1982) to incorporate other diffusion effects. They
applied the epidemic model to the adoption, and then rejection, of black and
white TV sets due to the rise of colour TV. However epidemic type models are
not generally used in technological diffusion as the models are deemed too
cumbersome, or have too many parameters for the limited available data (Mahajan et.al., 1990, p.13).

\subsection{Aims of Church Growth Modeling}

\label{Aims of Church}

What should such models achieve? Clearly the situation in a local church has
too many variable factors to allow for accurate prediction of numbers. Even
at the global level of a particular country parameters can change
unpredictably. It is tempting to fit models to actual data, however the
complexity of the underlying effects may make identification of the
processes difficult. For example, attempts to interpret USA church growth
data in terms of revival or lifecycle effects have proved controversial,
(see Miller and Nakumara (1996)  and references therein) and
demonstrates the difficulties involved.

Nevertheless mathematical models will provide useful information. Four
important results of modeling are:

\begin{enumerate}
\item \textbf{Principles.} Mathematical models can provide principles rather than numbers. An
example of this is seen in the predator-prey model originally developed
by Lokta and Volterra. There are few cases where the model fits well with
real data, but it does furnish the principle, called Volterra's principle,
that moderate harvesting across both species will cause the numbers of the
prey species to rise (Braun, 1975). The principle is well observed in
the fishing industry and in crop-spraying programmes, without actual data
being fitted to the model.

\item  \textbf{Understanding.} A model can help in the understanding of the dynamical process, which
can lead to a theoretical assessment of strategies.

\item \textbf{Data Gathering.} The model can help to decide what type of data should be gathered to
best measure a church's effectiveness.

\item \textbf{Explanation of Behavior.} The model can help explain why there is such a wide variation in the
speed and extent of church growth and decline. For example some growth is
slow and steady, whereas some, often associated with revivals, are fast.
Some revivals last many years, as in the 18th century Great Awakening, some
only for a year or so as in the 1858--9 revivals.
\end{enumerate}

There is much of topical interest in church growth such as: church planting
strategies; attempts to evaluate methods of evangelism; analysis of church
attendance statistics; and speculation whether the recent phenomena of the
``Toronto Blessing'' will result in a revival among the western Christian
church. The latter topic is particularly interesting because there is a
great reluctance to call the ``Toronto Blessing'' a revival, even among its
supporters, simply because there has not yet been a large number of
converts (Robinson, 1993; Wimber, 1994). As will be shown
later in this paper, epidemic type growth, so typical of a revival, can
have a very slow increase in numbers in the early stages. It is hoped that
some elementary mathematical analysis will shed some light on these types of
areas.

\subsection{Overview of Paper}

The main aims of this paper are:

\begin{enumerate}
\item  To show that mathematics can be used to model the dynamics of growth
in churches;

\item  To investigate the claim that conversion growth is proportional to
contact between unbelievers and active, or infected, believers -- called
enthusiasts. That is, the spread of religion is a form of interactive
diffusion;

\item  To investigate the claim that the enthusiastic, or recruitment, phase
of enthusiasts is limited in duration, after which time they become
effectively removed from the conversion process. That is, those who diffuse
the religious innovation do so only for a limited period. The removed are
called inactive believers.
\end{enumerate}

The epidemic model is constructed from its foundations in section \ref{Basic
Epidemic}, with some simple conclusions presented in section \ref{Results}.
The justification for applying the epidemic model to church growth, together
with its two claims (aims 2 and 3), is given in section \ref{The Simple
Church Growth}. The model is investigated for a number of typical church
growth situations, and compared with data from a past revival, in section
\ref{Numerical Solutions}. The differences between this revised paper and the original, Hayward (1999), are described in endnote 1.

\section{General Epidemic Model -- Construction}

\label{Basic Epidemic}

\subsection{General Assumptions}

\label{General Assumptions}

Although epidemic models are well understood (Anderson and May, 1987; Bailey, 1975), the development and results of the basic three compartment
model used in epidemic theory give important insight into its church growth
application. Thus a simplified version of the development is given here. A
fuller version can be found in Anderson and May (1987).

In the simplest model of the spread of an epidemic, three categories of
people are considered, represented by the variables:\ \medskip

\medskip
\begin{tabular}{ll}
$S$ & The number of susceptibles \\
$I$ & The number of infectives \\
$R$ & The number of people removed from the system after having had
the \\
& infection 
\end{tabular}

A glossary of all symbols relating to the epidemic and church growth models
is given in appendix A.

A susceptible, $S$, becomes infected through contact with an infective, $I$. Once
infected it is assumed that they are immediately able to infect others, even
if there are no symptoms. That is, the latent period of the infection is
negligible. People spend a certain length of time, $\tau$, in the
infected category. This is the period over which they can infect others.
It may be the entire infectious period of the disease, or less if isolation
takes place on or after the time symptoms show, or the disease is otherwise
detected. Once a person is removed from the infectious state, it is assumed
they are no longer able to infect anyone again or become infected again.
(Those in the ``removed'' category, $R$, may be removed because they have been
isolated, or have died, or are now immune to the infection.) Thus the
epidemic model is a compartment model for the three categories whose
numbers are: $S$, $I$ and $R$, referred to as stocks. This is represented
diagrammatically in figure 1 using the stock-flow  system dynamics notation (Sterman, 2000). The flows, labeled \emph{Infect} and \emph{Cure}, represent the rates of change between categories: $i$ and $c$ respectively.  It will be further assumed that the total
population $N=S(t)+I(t)+R(t)$ is constant.     Thus the system forms three differential equation  of the form (\ref{sf.1}--\ref{sf.3}).
       \begin{figure}[!ht]
          \begin{center}
   \includegraphics[height=1.8cm] {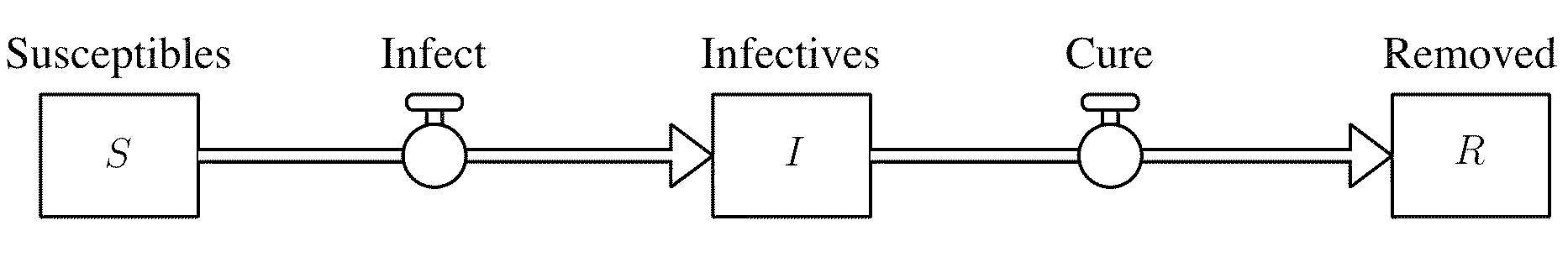}
       \end{center}
    \vspace{-15 pt}
    \caption{\small{Three compartment epidemic model.} } \label{fig1.fig}
 \end{figure}
\begin{eqnarray}
\dd S/\dd t &=& -i    \label{sf.1} \\     
\dd I/\dd t &=& i-c  \label{sf.2} \\
\dd R/\dd t &=& c\label{sf.3}         
\end{eqnarray}

\begin{figure}[!ht]
          \begin{center}
   \includegraphics[height=2.8cm] {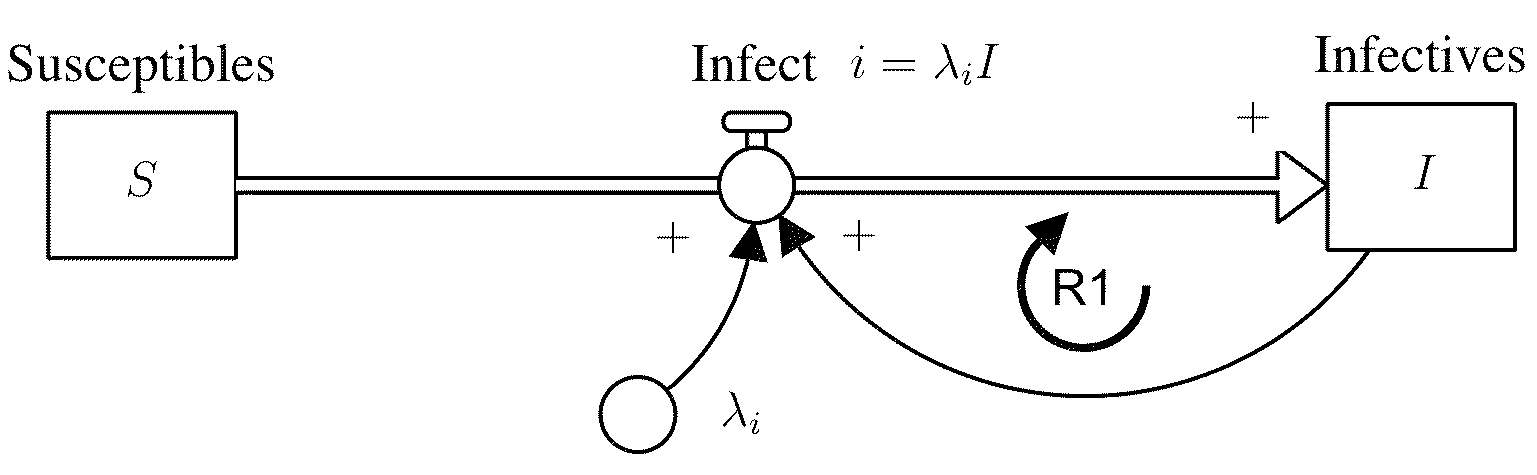}
       \end{center}
    \vspace{-15 pt}
    \caption{\small{Reinforcing feedback \small \textsf{R1} \normalsize on infectives in the epidemic model.} } \label{fig2.fig}
 \end{figure}

To obtain the transmission rates between the stocks further
assumptions are needed. The most fundamental, and most criticised
 of these is
homogeneous mixing (Anderson and May, 1987, p.65; Bartholomew, 1967, pp.215f, 247f). That is, the infectives are well mixed throughout the
susceptibles. (Other forms of contact can be considered, e.g. Anderson (1988), takes into account different degrees of contact amongst
susceptibles and infectives.) Homogeneous mixing  implies infectives will be equally likely to infect a
susceptible, thus the more infectives, the more infections per unit time, that is $dS/dt\propto - I$. Thus
\begin{equation}
dS/dt=  - \lambda_i I.
\label{e.1}
\end{equation}
where $\lambda_i$ is the per capita rate of infection. Thus, the infectives $I$ exert a force on the susceptibles as an increase in $I$ will cause an acceleration in the change of $S$. This force is illustrated by the connector from $I$ to \emph{Infect}  in the stock flow diagram, figure \ref{fig2.fig}.
A connector is a causal connection between two variables. Thus there is a feedback force  from $I$ to itself. This feedback is reinforcing, labeled \small \textsf{R1}\normalsize, as the more infectives, the more suscpetibles are infected, thus more are added to infectives. In the absence of other forces \small \textsf{R1} \normalsize gives exponential growth in $I$.

Different assumptions for how a disease is spread will give different forms
for $\lambda_i$. It is usual to make two assumptions. The \textbf{first assumption} is that the number of susceptibles infected by an infective during the whole of their infectious period, $n_S$, is proportional to the length of that period $n_S \propto \tau$. That is, the longer a person is infectious, the more people they infect. This is true for most diseases, although not for
people with a limited number of contacts such as non-promiscuous people with
sexually transmitted diseases (STDs). 

 It follows that  $n_S = \lambda_i \tau$, implying that the per capita rate of infection is independent of the infectious period, that is the infected person is regularly contacting new susceptibles. Thus 
 \[
 \lambda_i  = \frac{n_S}{\tau}
 \]

A further consequence of homogeneous mixing is that some of an infective's contacts will be with other infectives, in proportion to their fraction within society. Thus $n_S$ depends on the fraction of susceptibles in the total population, the probability of one infective contacting a susceptible $p_S=S/N$. Thus  $n_S  \propto p_S$, $n_S = n_N p_S = n_N S/N$, where $n_N$ is the number of susceptibles that would be infected by one infective, during the whole of their infectious period, given that the whole population is susceptible\endnote{$n_N$ is an idealized quantity as a population with all susceptibles contains no infectives to start the spread of a disease. The meaning of $n_N$ is defined in the limit as the number of infectives tends to zero (with  no removed). It measures the infectiousness of a disease.}.

\begin{figure}[!ht]
          \begin{center}
   \includegraphics[height=4.5cm] {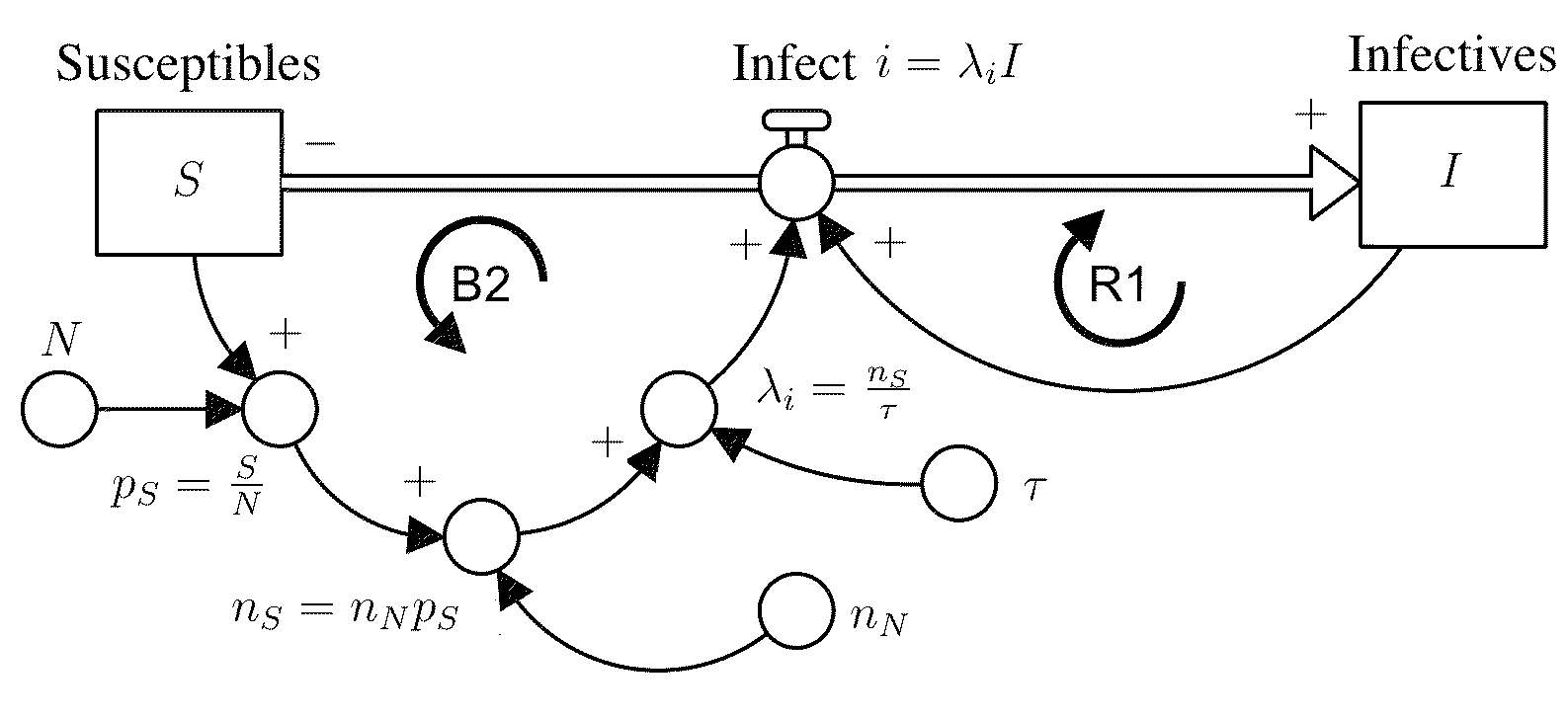}
       \end{center}
    \vspace{-15 pt}
    \caption{\small{Balancing feedback \small\textsf{B2} \normalsize on susceptibles in the epidemic model.} } \label{fig3.fig}
 \end{figure}

Thus the susceptible numbers $S$ exert a feedback force on itself, \small \textsf{B2}\normalsize, figure \ref{fig3.fig}. This feedback is balancing as there is one negative link in the loop. Thus, as $S$ declines, the probability of contact $p_S$ gets less ($+$ indicates a same way causal change), the various infection rates go down, thus less people are subtracted from $S$, slowing its decline. $S$ also exerts a force on $I$, opposing that from $I$, loop \small\textsf{R1}\normalsize, thus slowing the numbers being added to $I$ through infection. 

\begin{figure}[!ht]
          \begin{center}
   \includegraphics[height=4.5cm] {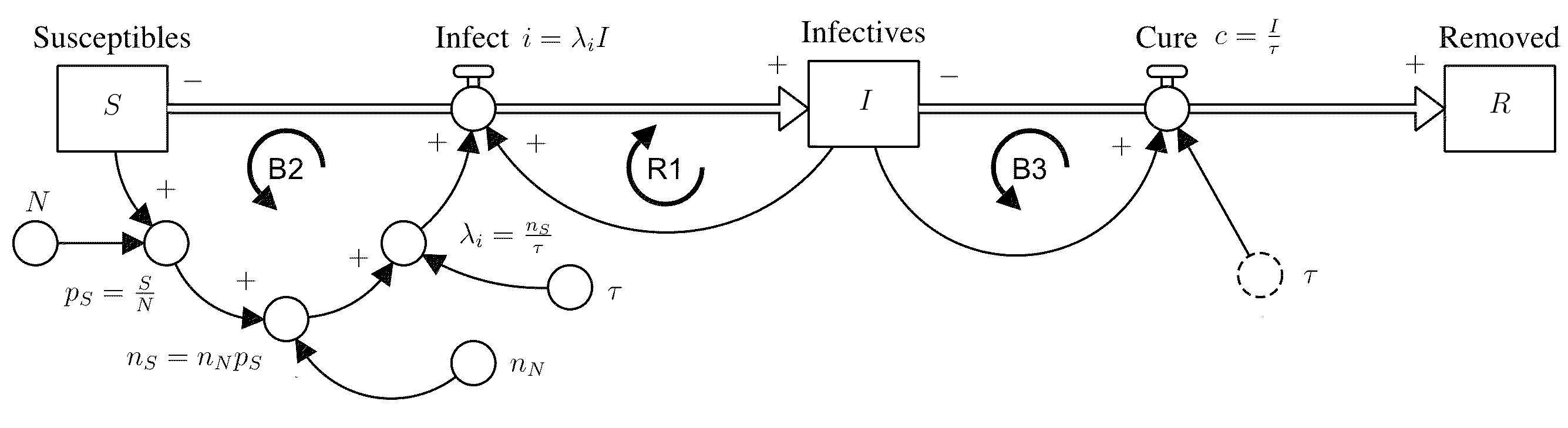}
       \end{center}
    \vspace{-15 pt}
    \caption{\small{Full epidemic model with three feedback loops:  \small\textsf{R1}\normalsize,  \small\textsf{B2} \normalsize and  \small\textsf{B3} \normalsize.} } \label{fig4.fig}
 \end{figure}

The rate of loss from the infectives, the cure rate, is proportional to their numbers, with a loss rate given by $1/\tau$. Thus the flow from $I$ to $R$ is a balancing loop, \small \textsf{B3}\normalsize, acting to oppose the increase in $I$ numbers, figure \ref{fig4.fig}. The more infectives, the more are cured, thus less infectives. Thus using equations \ref{sf.1}--\ref{sf.3},  with the equations in figure \ref{fig4.fig} the system dynamics model reduces to the differential equations (\ref{e.1}--\ref{e.3}).

\begin{eqnarray}
\frac{\dd S}{\dd t} &=& -\frac{n_N}{\tau N}SI    \label{e.1} \\     
\frac{\dd I}{\dd t} &=& \frac{n_N}{\tau N}SI -\frac{I}{\tau}  \label{e.2} \\
\frac{\dd R}{\dd t} &=& \frac{I}{\tau}\label{e.3}         
\end{eqnarray}



\subsection{Crowd Model}

\label{Crowd Model}

The \textbf{second assumption}, which will cover most infections, is that $n_N
\propto N$, the more people in the population, the more contacts an
infective has. Thus, a larger population will lead to more personal contacts, that is a network space of greater density.  This is typically what happens if a person with a cold
enters a class of students, the larger the class, the more students are
likely to get the cold in a given time period. It is also true if a larger
number of infectives enter a larger population, such as a university or a
town, in such a way that they are homogeneously mixed, an assumption already
made. This model is sometimes called the crowd model (Open University, 1988), also known as mass action (Hethcote, 1994), or density dependent (McCallum et. al., 2001). 

Thus, the form of $n_N$ is:
\begin{equation}
n_N=\beta \tau N  \label{e.4}
\end{equation}
where $\beta $ is a constant. Equation \ref{e.4} with equations \ref{e.1} to
\ref{e.3} give the classic equations for a general epidemic as given by
Bailey (1975):
\begin{eqnarray}
\frac{dS}{dt} &=&-\beta SI  \label{e.5} \\
\frac{dI}{dt} &=&\beta SI-\gamma I  \label{e.6} \\
\frac{dR}{dt} &=&\gamma I  \label{e.7}
\end{eqnarray}
where $\gamma = 1/\tau$ is the cure rate. 

The rate of transfer from $S$ to $I$ is given by the product
of the two population numbers, the ``mass action'' principle, originally
proposed for epidemics by Hamer (1906). This type of
non-linear behaviour in a compartment model is characteristic of epidemic
and many ecological models.

\subsection{Fixed Contacts Model}

\label{Fixed Contacts}

Clearly this second assumption is not true in the very early stages of an
epidemic where a lone infective in a very large population will not have
more contacts in a given time if the population is larger. Neither is it
true for STDs where there is usually a fixed number of different contacts
between an infective and other people in a given time period regardless of
the size of the susceptible population. Thus, increasing the population size will not change the number infected by one infective if the ratio of infectives to susceptibles stays the same; the system scales with population size. In these cases the second assumption
for $n_N$ is that it is independent of population size.  This has assumed that an infective
does not deliberately seek out susceptibles, thus their constant number of
contacts are shared between susceptibles, infectives, and the removed.
Thus the equations for the fixed contact model are given by 
\ref{e.1} to \ref{e.3}, where $n_N$ is constant. These are the general epidemic
equations familiar in the study of the spread of STDs and HIV/Aids (Anderson (1988). The model is also known as standard incidence, and frequency dependent (McCallum, 2001).

For most infectious diseases the crowd model is usually suitable, although
the truth often lies somewhere between the two models, the number of
contacts an infective has depends on $S$ but not linearly. Various
attempts have been made to construct a more sophisticated model of the
transfer from $S$ to $I$ (May and Anderson, 1985), however the two models are
sufficient to derive important epidemiological principles.

Of course, if the total population $N$ remains constant, the mathematical
results of the two models will be identical since all that has happened is
that the transfer constant $\beta $ has been redefined. However for
infections that are spread over long time periods, where there may be
births, natural deaths and migration, the two types of models will be
mathematically different. Again this will need careful consideration for
applications to church growth.


\section{General Epidemic Model -- Results}

\label{Results}

\subsection{Threshold of the Epidemic}

\label{Threshold}

Perhaps the most important result is that of the ``threshold'' of the
epidemic, first derived by Kermack and McKendrick (1927).
Technically an epidemic is deemed to occur if the rate at which people
become infected increases. This occurs when $\dot{I} >0$, which from (\ref{e.6}) gives a minimum condition for an epidemic as:
\begin{equation}
S_{0}>\frac{\gamma }{\beta }\triangleq \rho  \label{e.12}
\end{equation}
where $\rho $ is referred to as the threshold of the epidemic. This can be
demonstrated by illustrating the phase paths of $I$ against $S$ (figure 5).
Thus epidemics are more likely to occur in large concentrations of
susceptibles, a fact borne out by the prevalence of epidemics in large
cities and the general lack of epidemics among wild animals where large
concentrations are unusual. Epidemics are also more likely if the contact
rate between susceptibles and infectives ($\beta $) is higher, and if
the duration of the infectious stage ($\tau=1/\gamma $) is longer, both common
sense results. This principle has been demonstrated for numerous real
cases, one example is that of the Eyam plague of 1665--6 given by Raggett (1982).

       \begin{figure}[!ht]
          \begin{center}
   \includegraphics[height=7.0cm] {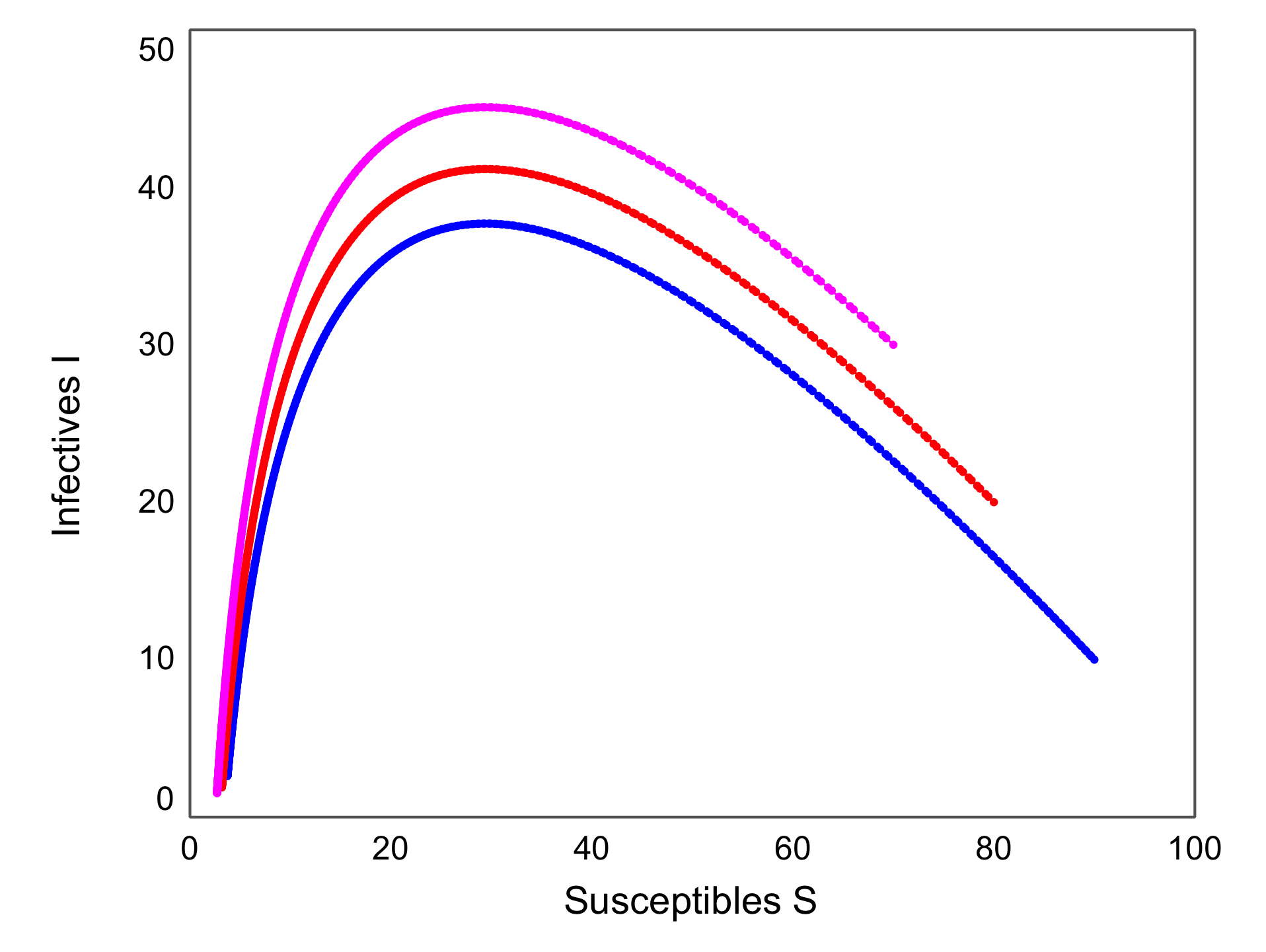}
       \end{center}
    \vspace{-25 pt}
    \caption{\small{Typical phase paths of the general epidemic.} } \label{fig5.fig}
 \end{figure}

Significantly, the value of $I_{0}$ does not influence the likelihood of an
epidemic (unless it is zero!). If the initial number of infectives is
smaller, the epidemic may take longer to start, but it will occur.

\subsection{Early Stages of the Epidemic}

\label{Early Stages}

In the early stages of an epidemic, when the number of infectives and
removed are very small compared with the total population ($S\simeq {}N$),
the growth in the number of infectives is exponential. Equation \ref{e.6}
becomes:
\begin{equation}
\frac{dI}{dt}=(\beta N-\gamma )I  \label{e.13}
\end{equation}
giving a doubling time for the early stages of an epidemic as:
\begin{equation}
t_{d}=\frac{\ln (2)}{(\beta N-\gamma )}  \label{e.14}
\end{equation}
This result is particularly useful in estimating $\beta $ if data is
available for those stages, for example Anderson (1988),  and May and Anderson (1987)   apply it to the spread of HIV. 

\subsection{End of an Epidemic -- Lack of Infectives}

\label{End of Epidemic}

Solving equations \ref{e.5}--\ref{e.7} in the static case gives $(S,0,N-S)$
as the equilibrium points for any $S$. Thus, on the $I-S$ plane, the
 $S$ axis for $S<\rho$ is in stable equilibrium, all solutions ending up at some
point in this region. Thus, some susceptibles remain at the end of the epidemic. What
determines its value?

There is no analytic solution of equations \ref{e.5}--\ref{e.7} for $S$, $I$
and $R$ in terms of $t$, however $I$ can be expressed in terms of $S$ and
their initial values by dividing (\ref{e.6}) by (\ref{e.5}):
\[
I(S)=I_{0}+S_{0}-S+\rho \ln \left[ \frac{S}{S_{0}}\right]
\]

The spread of the infection is over when $I$ becomes $0$, the number of
susceptibles remaining, $S_{\infty },$ can be calculated from the
non-linear equation:
\begin{equation}
S_{\infty }=I_{0}+S_{0}+\rho \ln \left[ \frac{S_{\infty }}{S_{0}}\right]
\label{e.15}
\end{equation}
$S_{\infty }$ is thus determined by the threshold and the initial values of $%
S$ and $I$. The epidemic ends not for lack of susceptibles but for lack of
infectives. The epidemic burns itself out before all susceptibles can catch
the disease because the infectives have fallen to insufficient numbers to
carry on the spread. This is due to there being insufficient infectives
initially for the number of initial susceptibles, given the threshold of the
epidemic. Increasing the number of initial infectives will always reduce
the susceptibles remaining as:
\begin{equation}
\frac{dS_{\infty }}{dI_{0}}=\left( 1-\frac{\rho }{S_{\infty }}\right) ^{-1}
\label{e.16}
\end{equation}
is always negative. Mathematically there is no value of $I_{0}$ that will
make $S_{\infty }$ zero, however there may be cases where it could be made
very close.

\subsection{Small Epidemics}

For a small epidemic Kermack and McKendrick (1927)  derived
a simpler result from equation \ref{e.16} namely: \textit{the number of
susceptibles falls to a value as far below the threshold as it started above}%
. This is referred to as the threshold theorem. Thus:

\begin{equation}
S_{0}-\rho \simeq \rho -S_{\infty }  \label{e.17}
\end{equation}

A small epidemic is one where the initial number of susceptibles is not far
above the threshold value, and the initial number of infectives is small.
However as Raggett (1982)  showed the theorem is not far from
the truth even for a fairly large epidemic such as the Eyam plague.

\subsection{Unlimited Infectious Period}

When the duration of the infectious period, $\tau$, becomes very long compared
to other timescales it can be treated as effectively unlimited and $\gamma
\rightarrow 0 $. In this case the system reduces to two equations in $S$
and $I$ only with $N=S+I$ a constant. Equation \ref{e.6} becomes the
logistic equation:

\[
\frac{dI}{dt}=\beta I(N-I)
\]
Thus the epidemic eventually spreads through whole susceptible population,
as in the standard models of cultural diffusion. This is sometimes referred
to as the simple epidemic model.

\section{The Limited Enthusiasm  Church Growth Model}

\label{The Simple Church Growth}

\subsection{Use of the General Epidemic Model}

\label{Use of Basic}

\begin{enumerate}
\item [a.]  The general epidemic model will be used as the initial model to
investigate the dynamics of how a church grows. The justification
for this is as follows: churches grow because people undergo a
process -- conversion -- which results in observable changes in a
person, such as church attendance, enthusiasm for the new faith,
adoption of a new moral code with its behavioural changes. The
rigid adherence to a distinct lifestyle has been recognised as an
important feature of growing churches (Kelley, 1986). Thus a
convert can be easily distinguished from an unbeliever just as a
person with an infection can be distinguished from a susceptible.
(The use of terms such as unbeliever and convert in the limited enthusiasm model are explained in appendix B.). Further,
Hadaway (1993a)  notes that enthusiasm among church
members has a significant effect on attendance, with effects
waning as enthusiasm wanes. 

\item [b.]  Most conversions occur because of a contact between an active believer
and an unbeliever, often via an interpersonal bond (Stark and Bainbridge, 1985, pp.309, 355). This active believer will be  called an \emph{enthusiast}\endnote{The name enthusiast was originally a derogatory term applied to people taken up with religion. In particular it was a nickname applied the the first Methodists in the 18th century who were instrumental in the conversion of others to the faith.}. The enthusiast may ``lead someone to Christ'' -- the
conventional expression used when a believer is instrumental in another
person's conversion. However the contact may simply be that an enthusiast takes the person to a church meeting or evangelistic campaign,
subsequently leading to a conversion at the hands of others. The growth in
the church is proportional to the contacts between an enthusiast and
unbelievers, just as the spread of an infectious disease is proportional
to the number of contacts between infectives and susceptibles. Hadaway (1993b)  notes that evangelism is an important predictor of church
growth. It is this contact process that slows down the growth into a
logistic behaviour. Without this process growth within a fixed size
population becomes unrealistically exponential as Stark and Bainbridge (1985, p.349) note in their example of cult growth.

\item[c.]  Not all people in the church are responsible for spreading the faith, that is not all are enthusiasts, using this paper's definition.
Indeed in most churches only a small proportion of believers are involved in
passing on their beliefs. For example, even in a highly successful ``Cell''
Church, 65\% of the membership being actively involved in the conversion
process is deemed a very high figure (Neighbour, 1990), no doubt a key
factor in their growth. For conventional churches the figure is more likely
to be less then 10\%.

\item[ ]  Thus, as well as enthusiasts (or ``infected'') believers, there are also
church members removed from most of the growth process. These are 
similar to the removed category in an infectious disease. Often it
is the new converts who are most enthusiastic about spreading the
faith, and who have the most non-Christian
contacts (Stark and Bainbridge, 1985, p.363). Thus, as a first approximation,
it is assumed that all new converts go through an initial phase of
enthusiasm where they are highly active in spreading the faith,
but, after a period of time, lapse into a less active role in
evangelism. Although the number of converts brought about by those
in the ``removed'' category will not be zero, it is assumed that
the number is very small compared to those from the infectives and
thus it can be ignored. This reduction of effectiveness is part of
the process of secularistion that many new churches undergo as
they becoming more accommodating to the surrounding society
 (Stark and Bainbridge, 1985, p.100, ch.19). However it will be too
simplistic to think of the removed as ``secularised'' believers.
It is purely the recruitment potential that is limited. The
reasons for this drop of enthusiasm are varied, but are usually
summed up in Wesley's Law of the decline of pure
religion (Kelley, 1986, p.55). Essentially, the ``law'' says that
taking up a new religion produces benefits, spiritual or material,
perhaps in the form of new friends, or respect, thus making
missionary zeal
more costly to engage in. It becomes easier to be devoted to work \textit{within} the church, rather than \textit{without}. The enthusiasm can also be
the result of an experience whose effects decline after a short period.

\item[ ]  Sometimes it is not a lack of enthusiasm that causes a drop in a
believer's usefulness. After a while most new converts find that they have
exhausted their network of non-Christian contacts. Some will cease to be
part of that network as the new convert exchanges old friends for new
Christian ones in their church (Olson, 1989).

\item[ ]  It is this process of a limited recruitment period that causes a
church to run out of potential converts as eventually happened to the early
Christian church (Stark, 1996, pp.12--13). It prevents a church from
eventually taking over an entire population, leaving sections of the
population untouched regardless of birth and death effects.

\item [d.] Periods of revival within the church often behave in a similar
fashion to an epidemic: there is a period where it builds up; it reaches a
climax; and eventually it passes away. It may take place gradually or
suddenly (Lloyd-Jones, 1986, pp.105--106). Not all church growth is
like this, neither do all diseases spread like this; there are endemic
infections. However epidemics and revival church growth share these
dynamical features.
\end{enumerate}

A number of different processes can be identified as causes of growth and
decline in an individual church. Growth is usually divided into three
categories: \textit{biological} (those born to church members, who
themselves become members); \textit{conversion} (those who become members
having had no upbringing in the church); \textit{transfer in} (those who
move into one church having left another). All three have their opposite in
terms of decay: death, reversion and transfer out. In addition there are
those who having left the church are \textit{restored} back. These processes
are explained in Pointer (1987,  pp.19--22).

For the main part of this paper the timescale will be chosen so that births
and deaths can be ignored to a first approximation, that is there is no biological
growth or decay. Thus the static epidemic model will be explored and its
shortcomings pointed out, where appropriate.

Further to this, growth by transfer, so significant for individual
congregations, will be ignored as the model will be mainly applied to the
church as a whole rather than one small part of it.

Thus the limited enthusiasm church growth model is the general epidemic model given by
equations  \ref{e.1}--\ref{e.3} or \ref{e.5}--\ref{e.7},  depending on
the transmission mode and population numbers.

\subsection{Identification of Variables and Parameters}

\label{Identification of Variables }

Given that the general epidemic model is a suitable starting point
to analyse church growth, the variables are easily identified:
\begin{itemize}
\item The  susceptibles are those not in the church, the 
\textbf{unbelievers} with whom the church members have contact. Isolated
unbelievers are not part of the dynamics of growth.

\item  The infectives are the \textbf{enthusiasts}, or
``infected'' believers, within the church who are active in spreading the
faith, that is active in making contacts with unbelievers that lead to their
conversion.

\item  The removed are those in the church who have a
negligible role in making converts, the \textbf{inactive believers}.  
\end{itemize}
Let the numbers of unbelievers, enthusiasts and inactive believers be $U$, $A$ and $B$ respectively. Thus $A+B$ is the total number in the
church, referred to as church members or believers. $N = U+A+B$ is the total
population involved in the dynamics of the growing church, which is assumed constant
in the short term. 

The parameter $n_N$, equation (\ref{e.2}) and figure \ref{fig3.fig}, is a measure of the effective contacts
between enthusiasts and unbelievers. This parameter represents the number of converts (not contacts) one enthusiast (infected
believer) is responsible for during the whole of their infectious
period before they drop down to the lower level of activity
characterised by the removed category. It is renamed the \textbf{conversion potential}, $C_p$, to reflect the change of application.   $C_p $ may be small
because the church is in high tension with society permitting very
few contacts  (Stark and Bainbridge, 1985, p.136). However $C_p $ could
also be small because the church is so much like society it has
nothing to offer, and although it has many contacts, few are
effective (Kelley, 1986, ch.6). For example the exclusive nature
of early Christianity made it far more effective than the moderate
pagan religions, even though the pagans had more actual
contacts (Stark, 1996, p.204f).

The parameter $\tau (=1/\gamma )$ is the length of time a believer remains infected or
enthusiastic. It is renamed the \textbf{duration a believer is active}, $\tau_a$.

These parameters, $C_p$ and $\tau_a$, may depend on a large number of sociological
factors in the surrounding society, or in church (Hoge and Roozen, 1979)  as well
as psychological and spiritual factors in the believer and unbeliever (Wagner, 1987). However it is assumed that for large enough numbers they
remain constant over a period of time, apart for the possible dependency of $C_p$ (i.e. $n_N$) if the crowd model of transmission is used. A change in one of the underlying
factors will result in a change in one these parameters and hence in the
dynamics of the church growth. This is discussed in section \ref{Medium Term
Revival - Global View}. The dynamical model should be relevant for
situations where growth depends on social context, or institutional factors,
or both.

\subsection{Identification of Transmission Mechanism}

\label{Identification of Transmission}

In section \ref{Basic Epidemic} two models for an epidemic were identified
depending on how $n_N$ depends on the population number: the
crowd model and the fixed contacts model. To decide which model is more
appropriate the transmission mechanism between an enthusiast and an
unbeliever needs to be identified. The key question is: if the population of
unbelievers is increased will each enthusiast  convert more people because they can contact more unbelievers? If the population is small then the answer is generally
``yes'' and thus the crowd model is more appropriate. However, in a larger population the answer is `'no''. Consider the following
transmission mechanisms:

\begin{enumerate}
\item [(i)]  The enthusiasts are engaged in a systematic program of
evangelism such as door-to-door work. In a small population then the larger the population the more
people will get visited -- thus the more contacts will be made, i.e. the
crowd model. However, there is a maximum number of people an individual can contact in one day, and a limit to the resources a church can provide for evangelism. Thus, once the population gets to a certain size, a larger population will \underline{not} lead to any additional contacts, therefore the fixed
contacts model is more suitable.

\item [(ii)]  The enthusiasts evangelise through their network of contacts.
Such social networks are seen as a major means of spreading the Christian
faith (Olson, 1989; Stark and Bainbridge, 1985, p.312f). This network is
unlikely to be larger if the population increases -- there are only so many
friends and acquaintances a person can hold down; this would imply the fixed contact model. 

It could be argued that  in a larger
population this network is often more changeable over time -- this increases
the number of contacts, and the number of people two or more believers have
in common in their network will be smaller, thus the number of global
contacts for the church is bigger. This would support  the crowd model. However, the timescale over which a network would significantly change is much longer than the duration of the enthusiastic period, unless it were a very mobile population, thus the mechanism is unlikely.

\item [(iii)] The enthusiasts are those caught up in a revival. In this
case, in their enthusiasm, they make contact with many people outside of
their normal friendship network. Indeed people whom the enthusiasts
have never met may seek them out simply because of news about them, and their
behaviour, has reached those people (Edwards, 1990, pp.90--91). This could 
 lead to an increased number of contacts in a larger population, but again there will be physical limits to the influence one individual can exert, unless they are using mass media. This latter mechanism would be better modelled separately to spread by contact, thus the fixed contacts model should still hold.
\end{enumerate}

It is anticipated that the limited enthusiasm model will be applied to a church consisting of many congregations in a region or nation. With such a population size the fixed contacts model should suffice, especially as the church is usually a minority, thus is unlikely to be short of contacts with unbelievers. If the model were to be applied to a single congregation in a village where population size is below the maximum number of contacts an individual could hold, then the crowd model would be more appropriate. The fixed contacts model will be assumed for the remainder of this paper.

The value of the enthusiastic period, $\tau_a$, will vary according to the mechanism. In some revivals it
can simply be a matter of months before the enthusiastic phase passes --
short term growth. In a programme of evangelism it is more likely to be
around two or more years -- medium term growth. It is conceivable that the
enthusiastic phase could last many years leading to long term growth,
however the general epidemic model is unsuitable as births and deaths have
been excluded.

\subsection{Inactive Converts}
With most infectious diseases each person who becomes infected are themselves infectious. This may not necessarily be the case with the spread of belief. Thus it assumed that not all the new converts become enthusiasts, but become inactive straight away and remain so\endnote{The concept of inactive converts was introduced in Hayward (2002), from which this section is taken. It has subsequently become a central hypothesis in the limited enthusiasm model and its extensions, (Hayward, 2005).}. There are a number of reasons for this:
\begin{enumerate}
\item They may be naturally shy and unwilling to engage in any form of recruitment;
\item They may be a social isolate and have virtually no network of friends to influence;
\item They may be a secondary convert, the spouse or child of a primary convert, who has “converted” for social reasons. It was common practice in the early church for the pagan husbands of Christian women to “convert” to the church (Stark, 1996, pp.~111-115). Often such secondary converts have little real enthusiasm for the actual faith;
\item It is possible for people to be converted to the ethos of the church -- its services, customs, and morality -- without ever being converted to the truth of the faith. As such they may have little desire to see others converted. Their `'conversion'' has been a purely social phenomena rather than one of deep religious conviction. Nevertheless they are part of the church, albeit an inactive believer.
\end{enumerate}
Thus only a fraction of the converts will become enthusiasts, with the remaining converts becoming inactive believers without ever being active in conversion.

\subsection{Equations of the Limited Enthusiasm Model}
\label{Equations of the Limited Enthusiasm Model}
\begin{figure}[!ht]
          \begin{center}
   \includegraphics[height=7.3cm] {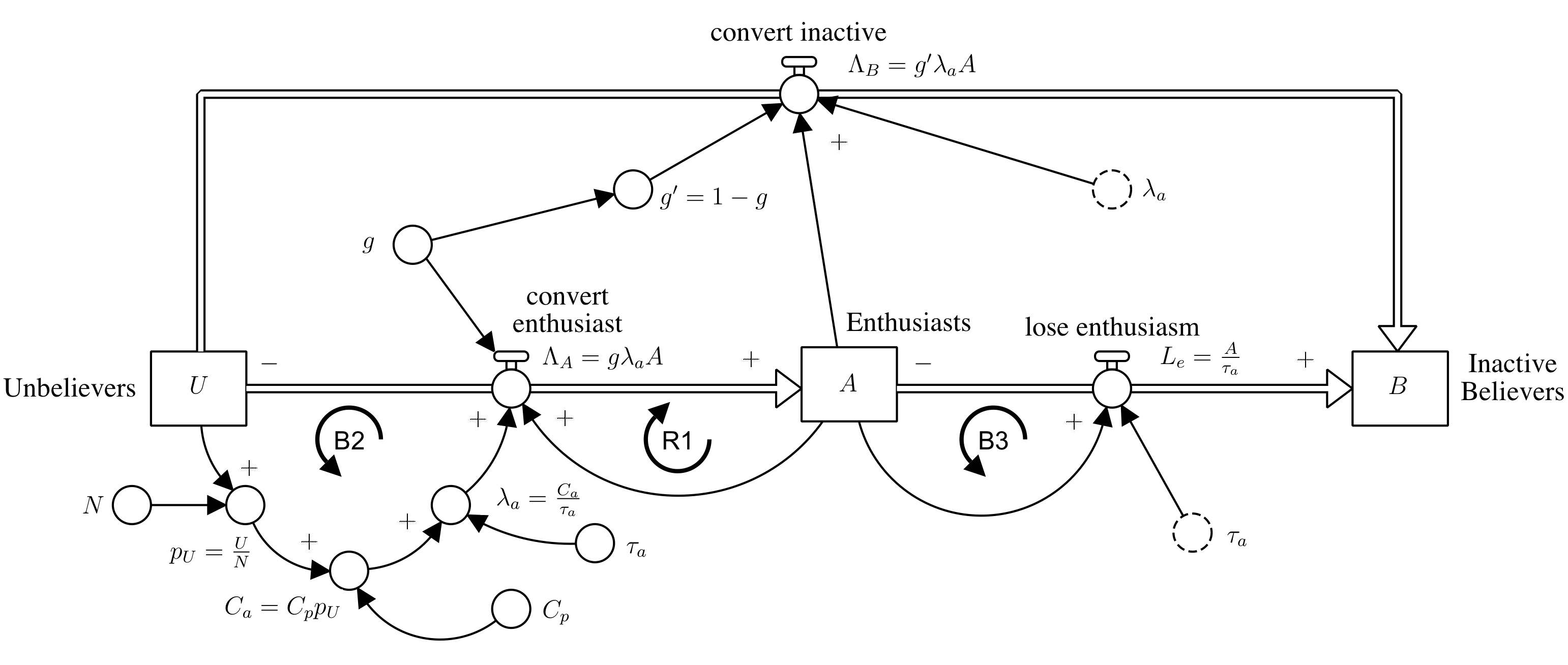}
       \end{center}
    \vspace{-15 pt}
    \caption{\small{Limited Enthusiasm Model of Church Growth. (Definitions see appendix A.)} } \label{fig6.fig}
 \end{figure}

The limited enthusiasm model can now be expressed in stock-flow form, figure \ref{fig6.fig},  using the fixed contacts model of figure \ref{fig4.fig} with the addition of inactive converts and the notation changes already discussed. A full description of each variable is given in the glossary, appendix A.  Inactive conversion is indicated by a flow directly from unbelievers to inactive believers. A fixed fraction, $g$, of converts become active, with $1-g$ becoming inactive immediately.

The stock-flow model reduces, through substitution, to the differential equations of the limited enthusiasm model (\ref{le.1}--\ref{le.3}):
\begin{eqnarray}
\frac{\dd U}{\dd t} &=& -\frac{C_p}{\tau_a N}UA    \label{le.1} \\     
\frac{\dd A}{\dd t} &=& \frac{gC_p}{\tau_a N}UA -\frac{A}{\tau_a}  \label{le.2} \\
\frac{\dd B}{\dd t} &=& \frac{(1-g)C_p}{\tau_a N}UA + \frac{A}{\tau_a}\label{le.3}         
\end{eqnarray}

The three feedback loops of figure \ref{fig6.fig} encapsulate the three central hypotheses of the limited enthusiasm model:
\begin{itemize}
\item [] \small \textsf{R1} \normalsize Enthusiasts generate enthusiasts through conversion;
\item [] \small \textsf{B2} \normalsize The generation of new converts is resisted through the depletion of the pool of unbelievers
\item [] \small \textsf{B3} \normalsize Enthusiasm is limited in duration. Enthusiasts eventually cease to make new converts and become inactive.
\end{itemize}
Each of these feedback loops acts like a force between stocks where  changes in one stock induces a deviation from uniform behaviour in another stock, or itself (Hayward, 2015; Hayward \& Roach, 2018). Thus, for example, \small \textsf{R1}\normalsize,  \small \textsf{B2} \normalsize and  \small \textsf{B3} \normalsize exert forces on the enthusiasts, $A$. These forces determine the pattern of behaviour in each dynamic variable (stock), and will  be used to help analyse the model (see section 5.8). 

An additional model assumption is that not all converts are enthusiasts. This is illustrated in the stock-flow diagram, figure  \ref{fig6.fig}, by the causal connections from $A$ and $U$ to the flow $\Lambda_B$. (The connection from $U$ is ``ghosted'' via $\lambda_a$.) Both form feedback loops, which for clarity have not been labeled, however they are dependent on each other. Thus the whole system has four independent feedback loops.

\subsection{Interpretation of Epidemic Model Results}

\label{Interpretation of Epidemic Model}

In section \ref{Results} four results were identified for the general
epidemic model. These can be applied to the limited enthusiasm model of church growth.

\begin{itemize}
\item[ ]  \textbf{Epidemic Threshold}. There is a threshold above which
significant church growth, or revival growth, will take place, the epidemic phase  (\ref{e.12}). In the church growth case the condition is given by $\dot{A}>0$, which from (\ref{le.2}) gives the condition for revival growth to take place as:
\begin{equation}
R_p \triangleq gC_p > R_{\mathrm{revival}} = \frac{1}{\bar{U}} \label{rev.1}
\end{equation}
where $\bar{U}$ is the fraction of unbelievers in society, $\bar{U}=U/N$, and $R_p$ is the reproduction potential, that is how many enthusiasts are produced by one enthusiast during their enthusiastic period. $R_{\mathrm{revival}} $ is the revival growth threshold that the reproduction potential must exceed for revival growth to take place.

Alternatively,   revival growth can be stated in terms of the initial number of unbelievers. The fraction of unbelievers initially must exceed the inverse of the reproduction potential
\begin{equation}
\bar{U}_0  > \bar{U}_{\mathrm{revival}} =  \frac{1}{R_p} \label{rev.2}
\end{equation}

Thus, growth is more likely to occur in large concentrations
of unbelievers for a given reproduction potential (\ref{rev.2}). Also, if enthusiasts reproduce more enthusiasts then revival is more likely for a given number of unbelievers (\ref{rev.1}) .  This agrees with common
sense, an important guideline in mathematical modeling. However the \emph{number}
of enthusiasts does not determine whether growth will take place or
not. A small church is as equally likely to see revival growth as a larger
one if their enthusiasts are equally effective; it will just take longer for the revival to get going and be spread
over a longer period of time. This will be investigated further in section
\ref{Numerical Solutions}.

\item[ ]  \textbf{Early Stages}. In the early stages a church grows
exponentially, equation \ref{e.13}. Such growth has been seen amongst South
American Protestant churches throughout this century, and among the
Pentecostal and New Church streams in the UK in recent years (Brierley, 1993). When the early phase is over, the growth usually slows down
in a logistic fashion.

\item[ ]  \textbf{End of Growth}. Growth eventually comes to a halt because
of a lack of infected believers. The church runs out of enthusiasts, because
their conversion rate is not sufficient among a falling number of
unbelievers. Growth does not end because there are no more unbelievers. The
history of revivals show that they stop long before all the people in a
population are converted or reached. However a church with more enthusiasts
at the beginning will see greater growth, all other things being equal, as
equation \ref{e.16} shows.

\item[ ]  \textbf{Threshold Theorem}. The number of converts made during a
period of growth will be approximately double the difference between the
number of unbelievers and the threshold.

\item[ ]  \textbf{Limited Enthusiasm}. If enthusiasm is not limited in
duration then religious belief would spread along the lines of classic social
diffusion and eventually cover the entire population. \emph{It is the limitation
of enthusiasm that prevents the whole susceptible population being
ultimately converted}. It is the thesis of this paper that church growth is limited by the limited period of effectiveness of enthusiasts.
\end{itemize}

An example is helpful to illustrate the last point.  Let the church be in a
population of say $50,000$. Assume the church is small, e.g. less than 50 people; initially composed entirely of enthusiasts. Thus $U_0 \approx 50,000$. As an average, let $10$ enthusiasts be
responsible for making $22$ converts during their enthusiastic period, of whom only $11$ become enthusiasts. Thus  $C_p=2.2$ and $g=0.5$ giving the reproduction potential as  $R_p=1.1$ from (\ref{rev.1}). The revival threshold fraction of unbelievers is then $ \bar{U}_{\mathrm{revival}} = 1/R_p = 0.909$, from (\ref{rev.2}). Thus the threshold value  is about $45,500$ unbelievers,  giving a
difference from the initial number of unbelievers of about $4,500$. Thus around $9,000$ converts are made, with the bulk of the population remaining unconverted. Limited enthusiasm has given limited growth. 

In the example a figure of 9,000 conversions in a population of 50,000 appears very high.  In reality, in a
typical British town of $50,000$ people, many churches will contain no such
enthusiasts. Thus the number of initial enthusiasts $I_{0}$ is very small,
and this growth would occur over a period of time much longer than the
lifetime of the individuals. The growth, therefore, has to be offset by
deaths. Thus a few churches see some growth, and the rest survive or die due
to biological and transfer effects alone.

Further, the churches may not be in effective contact with a significant
proportion of the population of 50,000 for reasons of geographic location,
class, race etc. Thus the actual growth is smaller, drawn from the church
members' circle of influence only.

\section{Model Results}

\label{Numerical Solutions}

\subsection{Numerical Solution}

\label{Scaled Equations}

The limited enthusiasm church growth model is a non--linear system without an analytical
solution in general. Thus, to investigate time scales for growth, and the
number converted, the differential equations are solved numerically using the system dynamics software Stella Architect, produced by ISEE systems. Results were also checked with the 
Runge-Kutta-Fehlberg method of order 3/4 (Burden and Faires, 1988) programmed in Ada 95.

\subsection{Increasing the Effectiveness of an Evangeliser}

\label{Increasing the Effectiveness}

One aim of evangelistic programs is to increase a believer's effective
witness. One approach is to train people to explain the gospel effectively.
Many methods are taught throughout the church, a number of which are
reviewed by Green  (1990, part 3).

The effective witness can also be improved by increasing the number of
contacts with unbelievers. Two models of church organisation that attempt to
achieve this are the Seeker Church model, pioneered by the Willow Creek
Community Church near Chicago (Robinson, 1995), and the Cell Church
model. Examples of the latter include the Yoido Full Gospel Church in Seoul,
Korea, and the underground church in China, both of which have seen huge
growth in recent years. Cell Church methods are explained by Neighbour (1990).

The idea behind both approaches, which can be employed together, is that,
all other things being equal, a believer who has been so trained will be
responsible for more conversions. Such people are candidates for being
treated as enthusiasts, and will be called evangelisers in this
paper. (The term evangelist has a more technical meaning within the
Christian church.)

Assume that the effectiveness of such a method is $\tau_a=1$
year, i.e. a believer loses their evangelistic impact one year after
conversion, on average. Assume also that the number of enthusiasts is
initially $5\%$ of the church, with $10\%$ of the total population in the
church, thus $\bar{U}_0=0.9$.  Thus, from (\ref{rev.2}), the threshold reproduction potential is $ R_{\mathrm{revival}} = 1.11$. Finally, assume that only $50\%$ of the converts become enthusiasts, $g=0.5$. Thus it requires a conversion potential of $C_p= 1.11/0.5 = 2.22$ for the number of converts to exceed the threshold for revival growth.

The equations can be solved with a variety of values of $C_p$ from $1.0$
up to $3.0$ converts per infective over that one year period.  The
percentage church growth over a fifty year period is shown in figure 3. Note
the effect is near exponential. (This effect was computed on arithmetic
arguments by Stark and Bainbridge (1985, p.355), although it was not explicitly
stated.) The benefits from doubling the effectiveness of an individual
believer is to more than double the growth rate of the church. This effect is especially marked near the threshold value of $C_p=2.2$. Thus increasing the effectiveness of evangelisers, an effectiveness they pass on to their converts, is a powerful strategy for church growth. This could be achieved through beginners courses such as the Alpha Course and Christianity Explored, where new converts on the course help on future courses, bringing along their friends. 

       \begin{figure}[!ht]
          \begin{center}
   \includegraphics[height=7.0cm] {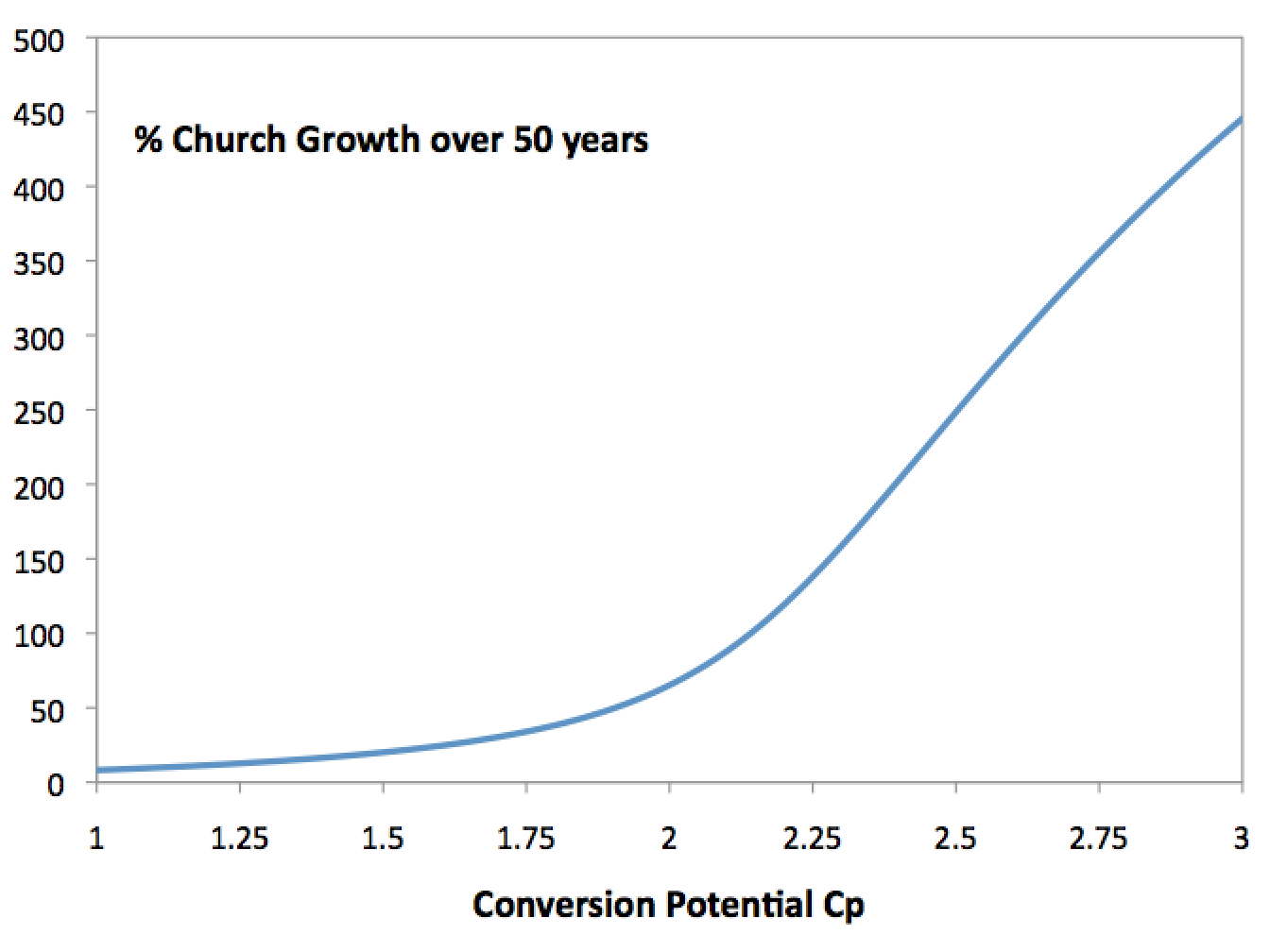}
       \end{center}
    \vspace{-25 pt}
    \caption{\small{Effect of changing conversion potential (individual effectiveness) $C_p$ on church growth after 50 years. $N=100$, $C_0=A_0+B_0= 10$, $A_0/C_0=0.05$, $\tau_a = 1$ year, and $g=0.5$.} } \label{fig7.fig}
 \end{figure}

\subsection{Increasing The Number of Evangelisers}

\label{Increasing The Number}
\begin{figure}[!ht]
          \begin{center}
   \includegraphics[height=7.0cm] {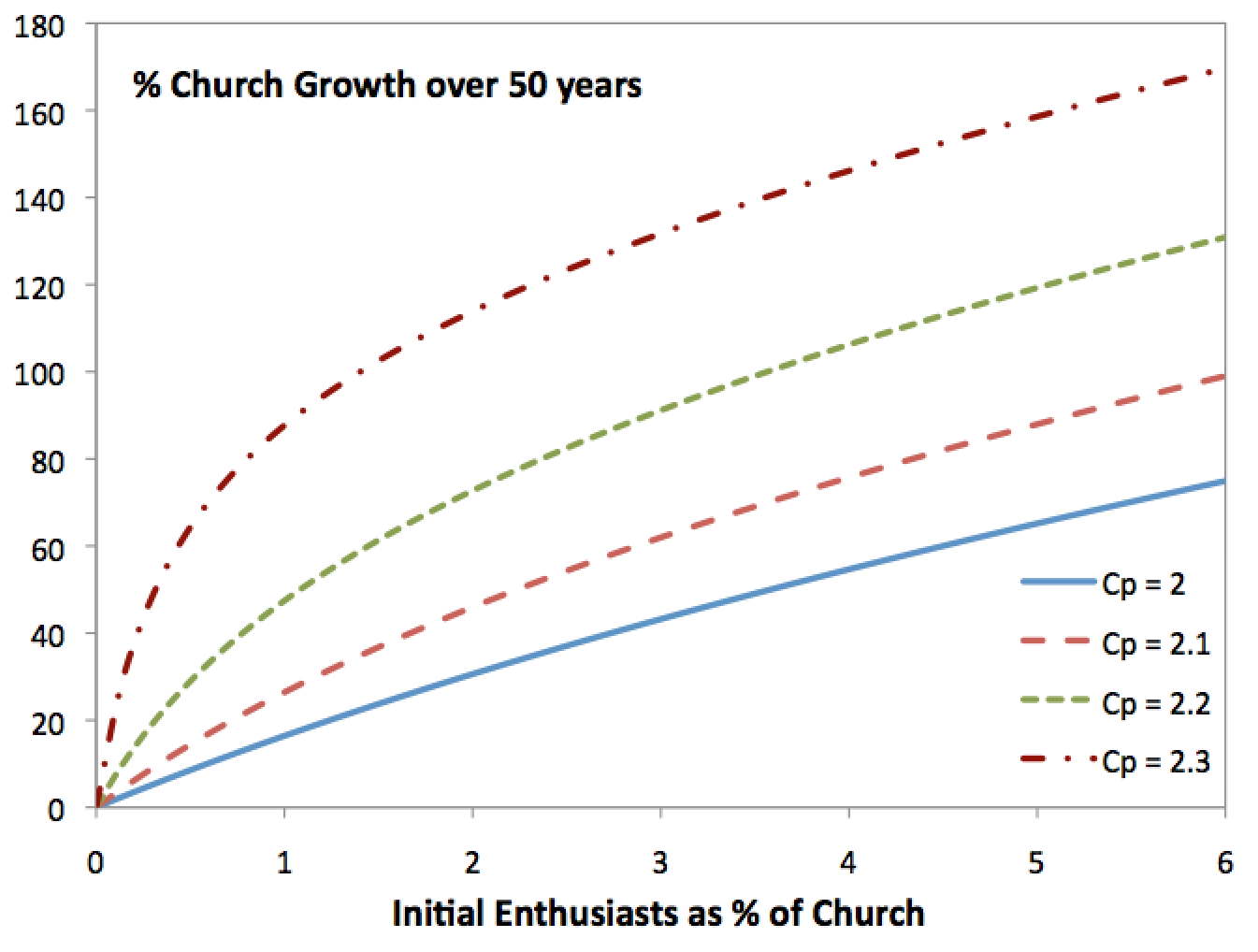}
       \end{center}
    \vspace{-20 pt}
    \caption{\small{Effect of changing the initial  percentage of evangelisers $A_0/C_0$ on church growth after 50 years. $N=100$, $C_0=A_0+B_0= 10$, $C_p=2, 2.1, 2.2, 2.3$, $\tau_a = 1$ year, and $g=0.5$.} } \label{fig8.fig}
 \end{figure}

Another aim of evangelistic programs is to increase the number of people
involved in evangelism. Keep $\tau_a$ at $1$ year, $g=0.5$, and choose four values $C_p$ ranging from just below the revival threshold to just above,  $C_p = 2, 2.1, 2.2, 2.3$.
If $A_0/C_0$ is now varied the percentage church growth responds in a linear
fashion for values of $C_p$ below the threshold, and slower than linear as $C_p$ is increased above the threshold, (figure \ref{fig8.fig}). Thus increasing the number of evangelisers does not have
the same impact as increasing an evangeliser's effectiveness.

To explain this result note that in the early stages the increase in the
number of infected believers is approximately exponential in time $
\propto A_{0}\exp(\eta t)$, where $\eta$ is proportional to the conversion potential. This expression is linear in $A_{0}$ but
exponential in $\eta$, thus growth is more sensitive to changes in
effectiveness than it is to the initial number of enthusiasts.

\subsection{Long-Term Revival}
In subsections \ref{Increasing the Effectiveness}--\ref{Increasing The Number} the church  took about 50 years to grow to near its final value. This type of revival is  described as long-term as it is similar to a human generation; that is  a period where a significant number of  births and deaths in the church have occurred. Additionally, a large number will have left the church during this period, called reversion. These features are not included in the basic version of the limited enthusiasm model presented here, but are covered in a subsequent publication (Hayward, 2005). Nevertheless, the basic model captures the effect of changes in the model parameters and the significance of the revival threshold.

Historical examples of long-term revivals include the first and second great awakenings in the USA (Weisberger, 1966; Gaustad, 1968; Cross, 2015); the rise of Methodism in the UK during $18^{\mathrm{th}}$--$19^{\mathrm{th}}$ centuries (Evans, 1985; Ryle, 1978; Edwards, 1990); The Isle of Skye Revivals, 1800--1860 (Taylor, 2003 ); the rise of world-wide Pentecostalism in the $20^{\mathrm{th}}$ century (Riss, 1988; Martin, 2001); and the rise of the late $20^{\mathrm{th}}$ century Charismatic Movement (Hunt, 2009). Indeed the rise of Christianity itself from the first to the fourth century could be viewed as a very long-term revival (Stark, 1996). The Christian church has grown through revival many times through its history, though those earlier in history become harder to specify accurately due to the relative lack of historic documents.

\subsection{Medium-Term Revival}

\label{Medium Term Revival - Global View}
Many revivals in the Christian church occur over periods shorter than a generation, about 10--20 years, where births, deaths and reversion have only minor effects. Such medium-term revivals 
invariably start among its members first (Lloyd-Jones, 1986, pp.99--101)
, with the ``fire'' being spread from believer to believer before it reaches
unbelievers. This is sometimes called a renewal phase of a revival.
Mathematically it requires a mass action type contact between $A$ and $B$ to
model the change of behaviour among inactive believers, a feature the limited enthusiasm
church growth model doesn't contain\endnote{Renewal was added in Hayward (2010).}. However the  model will give some
indication of the later stages of a revival when contact with unbelievers
becomes the dominant behaviour. Believers affected by a revival spread the
gospel with considerably increased enthusiasm. Such believers are candidates
for being enthusiasts, those ``infected'' by the revival.

Consider a global view, i.e. the whole of the Christian church in one
country. Keeping the church as $10\%$ of the whole population (about the UK
figure), revival growth will occur if the initial  threshold $R_{\mathrm{revival}} =  1/\bar{U}_0 $ is exceeded by the reproduction potential $R_p$, as given by equation
\ref{rev.1}. Thus $R_{\mathrm{revival}} =1/0.9=1.11$. Of course the church will grow if $R_p$ is less than this figure but it will not be revival type growth
with the number of enthusiasts increasing. 

Assume that $C_p=2.3$ and $g=0.5$, giving $R_p = 1.15$, just in excess of the threshold. Let  believers only be infected for a short
period of $\tau_a=0.1$years, to ensure the revival is medium term. Thus converts make a significant impact on
unbelievers for only a short period after their conversion.

\begin{figure}[!ht]
          \begin{center}
   \includegraphics[height=7.0cm] {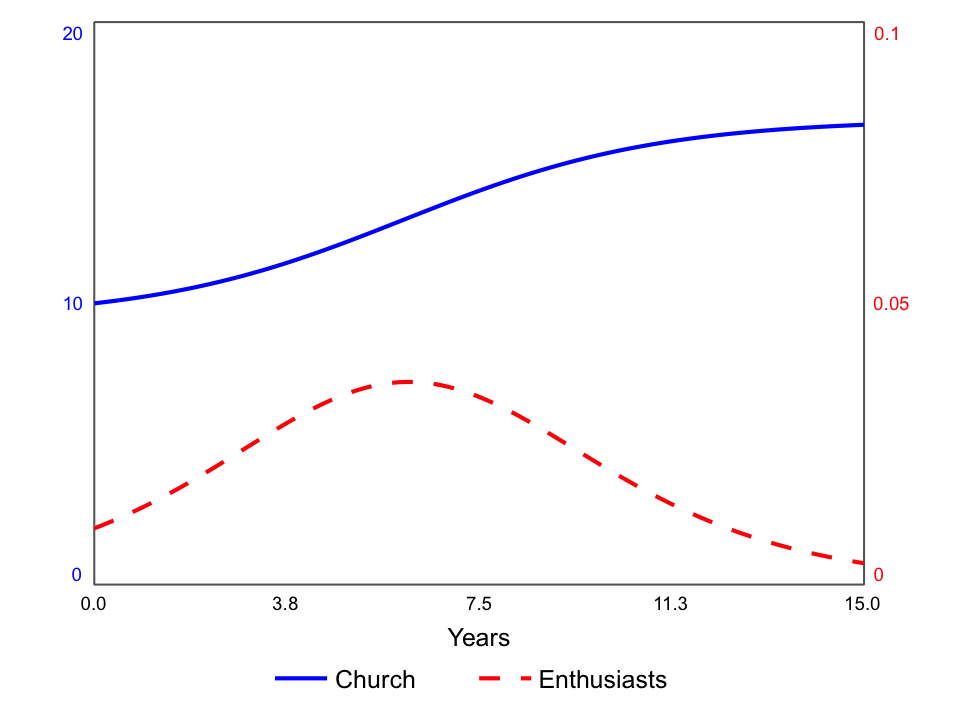}
       \end{center}
    \vspace{-25 pt}
    \caption{\small{Growth of nationwide church -- medium term revival. $N=100$, $C_0=A_0+B_0= 10$, $A_0/C_0 = 0.1\%$, $C_p= 2.3$, $\tau_a = 0.1$ year, and $g=0.5$} } \label{fig9.fig}
 \end{figure}

Typically revivals in a country start with a small number of infected
believers (Lloyd-Jones, 1986, pp.163--166). Let $A_0=.01$,  i.e. only
one in a thousand of the church are so affected. The resulting growth of
the church is given in figure \ref{fig9.fig}.

The church, initially $10\%$ of the population, increases to $16\%$ over $
15$ years. However, the start of the growth is slow with only $1\%$ of the
population added in the first 3 years. The bulk of the growth is in the
following $7$ years, which sees a further $4\%$ added. Thus a revival may
not be immediately noticeable in terms of a substantial increase in
numbers within the church. Bearing in mind that this follows an earlier
renewal phase, the time period before growth is noticed could be quite
lengthy.

This ``slow start'' behaviour typifies a medium to long term revival such as
the 18th century evangelical awakening in Britain. Although it started in
the 1730's the significant effects on church numbers did not occur until the
middle of the century with much of the increase in the latter half. One of
the reasons for the slowness of the revival is the low numbers of church
members within society as a whole, together with the low number of infected
believers initially. It is these conditions which prevailed in the 18th
century. By contrast the revivals during the 19th century in Britain and the
USA were faster, but the church was a much larger proportion of the
population. This will be investigated in section \ref{Short Term Global}.

Another significant result is that the revival is ending due to dynamical
effects dependent on its initial intensity, and the fact that a believer's
enthusiastic phase is limited. It is not ending due to any change in
spiritual conditions such as the revival work being hindered in some way.
Given that infected people are only effective for a fixed period then,
with a given number of susceptibles, only a certain number of conversions
become possible before the number of susceptibles an infected person is
likely to meet in that time period is too small to keep the revival going.
Of course the believer may still be involved in conversions after their
infectious period ceases, but this is at a much lower level and does not
give revival type growth.

The only way to increase the number of converts in a revival is to increase
the effectiveness of the enthusiast $C_p$, that is increase
the number of effective contacts between an enthusiast and an
unbeliever. This may be done by increasing the number of contacts, or by
unbelievers being more responsive to the gospel message. It is this latter
method that is deemed by the Christian Church to be a significant cause of a
revival taking place. Theologically, a revival is regarded as an ``act of
God'' which turns believers into effective witnesses and makes unbelievers
responsive to that witness 
(Lloyd-Jones, 1986, pp.50, 56--57, 106, 233--236).

Increasing $C_p$ to $1.2$ converts per person sees a larger but shorter
revival. The revival is over in about $10$ years with the church
increasing to about $22\%$ of the population. The church sees a $20\%$
increase in its numbers in  $3$ years, compared with a $10\%$ increase with
the lower figure for $C_p$. Thus the revival is noticed earlier.

Increasing the parameter $\tau_a$, the time period over which conversions take
place, slows the revival down but the numbers converted stay the same. Indeed $\tau_a$
could be removed from the equations by scaling the time $t$. Although having
a limited duration to the enthusiastic period limits the growth to a number
less than the whole population, its value does not effect the amount of
growth. Over longer periods, where births and deaths become significant, this
result will no longer apply.

Examples of medium term revivals are: the beginnings of the Methodist revival in the UK 1735--1760 (Evans, 1985; Ryle, 1978); the beginnings of the First Great Awakening in  Northampton Massachusetts (Edwards, 1984);  the Beddgelert Revival, Wales 1817--1822 (Davies, 2004); Nagaland Revival 1970s (Orr, 2000; Hattaway, 2006); the East African Revival, 1930s (Butler, 1976; Ward \& Wild-Wood, 2010); Azusa Street, Los Angeles, 1906--1915 (Bartleman, 1980); the  ``Toronto Blessing'' (Riss \& Riss, 1997; Poloma, 2003); and the Brownsville Revival 1995-2000, Pensacola (DeLoriea, 1997).

\subsection{Short-Term Revival}

\label{Short Term Global}

In some periods the church has occupied a much larger proportion of the
population. In Britain during the 19th century it accounted for nearly half
the population. Assuming the church is $50\%$ of the population, the
threshold for revival growth to occur is now higher, $R_{\mathrm{revival}}=2$, thus
more new enthusiasts per enthusiast are required for revival growth to occur.
Whether this is ``harder'' to achieve cannot be answered, there are too many
factors, however from a social point of view a church that is more
acceptable in a culture, because of its size and therefore influence, may
find it easier to make converts. Thus revival growth can occur albeit with a
larger value of $R_p$.

\begin{figure}[!ht]
          \begin{center}
   \includegraphics[height=7.0cm] {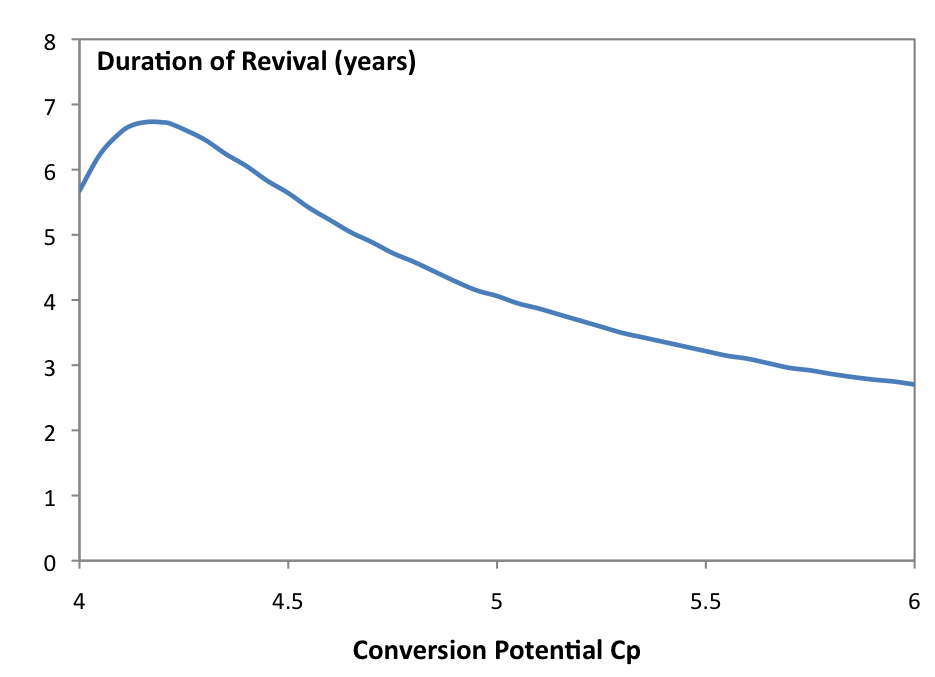}
       \end{center}
    \vspace{-20 pt}
    \caption{\small{Duration of short-term revival in a country. $N=100$, $C_0=A_0+B_0= 50$, $A_0/C_0 = 0.1\%$, $\tau_a = 0.1$ year, and $g=0.5$.} } \label{fig10.fig}
 \end{figure}

\begin{figure}[!ht]
          \begin{center}
   \includegraphics[height=7.0cm] {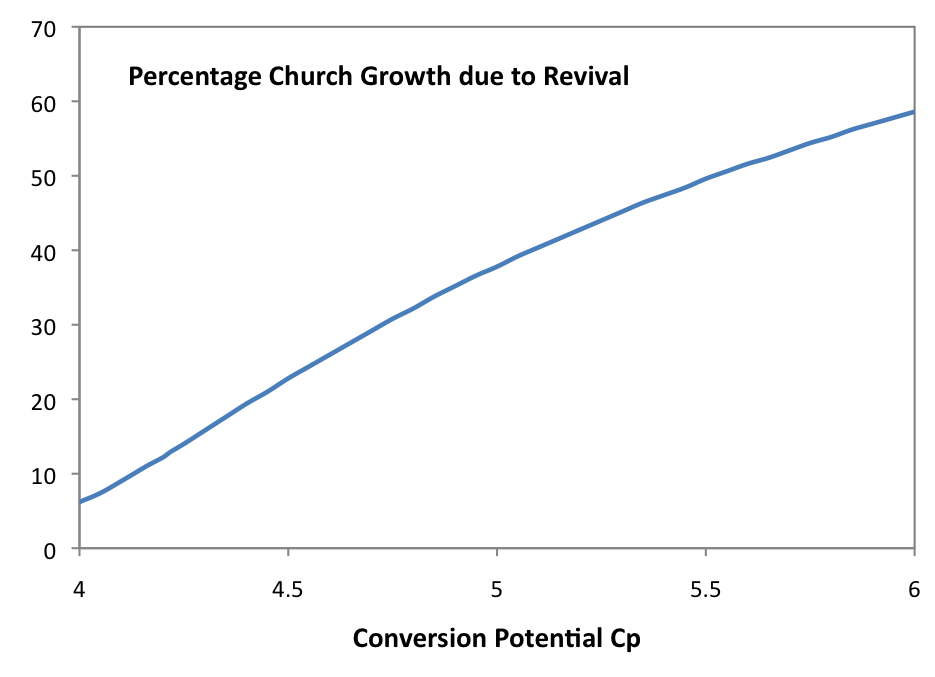}
       \end{center}
    \vspace{-25 pt}
    \caption{\small{Growth of a church in a country -- short term revival. $N=100$, $C_0=A_0+B_0= 50$, $A_0/C_0 = 0.1\%$, $\tau_a = 0.1$ year, and $g=0.5$.} } \label{fig11.fig}
 \end{figure}

For example, keep the population at 100 with the church now at 50, thus $\bar{U}=0.5$. Church occupies half of society. Keep the  initial number of enthusiasts at $0.1\%$ of the church, thus $A_{0}=0.05$. Keep the effective
period $\tau_a=0.1$ year, and $g=0.5$. The end of a revival is defined when the rate  of \emph{increase} in the church falls to a low level, specifically $0.1\%$ of the total population per year. For example for $C_p=4.1$ ($R_p = 2.05>R_{\mathrm{revival}}=2$) the revival ends after $6.5$
years and the church increases by $9\%$ of the population. Figures \ref{fig10.fig} and \ref{fig11.fig}
show the duration and increase in the church for values of $C_p$ from 4 up to $6$.

The larger the value of $C_p$, the faster the revival growth and the
larger the number of converts. The intensity of the revival is very
sensitive to the number of converts per person. Indeed for $C_p=5.3$
the cumulative conversion \textit{rate} is five times that when it is 4.1. Short-term
revivals can be become noticeable very quickly in their impact on society
around them! This contrasts quite significantly with the results of
section \ref{Medium Term Revival - Global View}. When the church is a larger
proportion of the population then, \emph{if} a revival occurs, and other conditions
remain the same, then it is likely to be faster and more intense. This fact
appears to be born out by history when the revivals of the 18th century in
a numerically weak church are contrasted with the much faster ones of the
following two centuries when the church was stronger.

Examples of short-term revivals are the Welsh revivals of 1859 and 1904 (Evans, 1967, 1969); the Wheaton College (Coleman \& Litfin, 1995); the Isle of Lewis, Scotland, 1939 and 1949 (Peckham \& Peckham, 2004);  Bario, Malaysia, 1973, 1975, 1979, 1984 (Bulan \& Bulan-Dorai, 2004); and East Anglia, UK, 1921 (Griffin, 1992).  Each of these revivals lasted about two years or less, thus very much at the shorter end of the time scale of figure \ref{fig10.fig}.

\subsection{Estimation of Parameters from Data}
Using real data from churches poses some considerable challenges. Most
churches keep a record of membership numbers, but the meaning of membership
varies. At one extreme there are very strict protestant churches, such as
the Methodists in the 18th century, where evidence of conversion must be
shown, and membership is discontinued if commitment is lacking. At the other
extreme some are very lenient, such as the Roman Catholic church, where all in
the religious community are members regardless of commitment. Attendance is
much higher than membership in the strict churches, and much lower in the
lenient churches. Thus membership is rarely a measure of religious attendance,
its relationship to attendance will differ between different churches and
over time. Until recently few churches recorded attendance, except at
untypical times such as Easter.

Revivals thrive on anecdotal evidence but reliable data is hard to come by.
Even the data collecting that does take place may prove unreliable during a
revival as individual churches are otherwise distracted by events in their
midst. It should also be noted that following a revival new religious groups
get formed for whom no data is available, thus an accurate picture is
impossible to achieve.

However an estimate can be attempted for the revival that took place in
Wales in 1904--5. Annual membership figures are available for all churches
apart from the Anglican church, for whom communicant figures are available (Williams, 1985). These are used as an estimate of membership. For 1904
the combined adult total for churches in Wales stood at $48.94\%$ of the
total Welsh adult population. Prior to this date the percentage had been
falling very slowly, until in 1904 it had risen 1 percentage point from
1903. By the end of 1905 the percentage of people in membership of Welsh
churches had risen to $53.43\%$ of the total adult population, where it has
had to be assumed that the Anglican church increased by the same percentage
as the other churches. (It changed its method of measuring communicant
numbers in 1905.) Anecdotal evidence would support this assumption. The
total adult population had also increased over the year by $2\%$ to stand at
1,446,447.

The revival started in October of 1904 in at least two separate geographical
locations with a very small number of people in each. A number of other
churches had become involved in the revival by the end of 1904. Assume that
about 1 in 1000 of church people had become enthusiasts for the revival by
the beginning of 1905, i.e. about $800$ people. The bulk of the converts
came in the next 12 months, so this will be taken as the duration of the
revival. For $g=0.5$ then the basic church growth model gives $R_p=2.008$ with a
duration $\tau=0.8$ weeks. Thus the actual reproduction of enthusiasts was $R_p\bar{U}=2.008\times0.5106 = 1.03$, i.e. each enthusiastic believer, on average, was
responsible for bringing slightly more than one new enthusiast into the churches in the space of a week. That is, 2.06 converts per enthusiast as $g=0.5$. If
the number of initial infectives at the beginning of 1905 is underestimated
then the value of $R_p$ remains about the same but the enthusiastic
period becomes shorter, that is, enthusiasts have to bring in new people
faster. The duration of the enthusiastic phase is very short, much shorter
than can be explained by any process of secularisation. Its shortness reflects that in a social phenomenon such as revival the infected may be infectious for a number of sporadic periods, whose sum is on average the figure for $\tau$. It is also noted that in this revival churches had daily meetings and a new convert would invite a friend to another meeting within a matter of days\endnote{The computation for the 1904--5 Welsh Revival in this revised paper is different from that presented in the original 1999 paper as only half the new converts have been considered enthusiasts, $g=0.5$. The 1999 paper did not have this feature, thus it automatically assumed $g=1$. There is no easy way to estimate the value of $g$ for this revival from the data used here.}.

Note that  with about half the population churchgoers, the reproduction potential $R_p$ would have to
be at least 2 enthusiasts per person for a revival to take place -- what ever
the time scale. It is very likely that this value for $R_p$ is only an
average and that a small number of enthusiasts were responsible for more than 2 enthusiasts over a longer period, with the bulk of the enthusiasts responsible for less over a shorter period. However, this would require a more sophisticated
model than the basic church growth one, and it is unlikely that any data
from this period could discover these variations.

\subsection{Social Forces and Revival Church Growth}
In the context of this paper a social force is defined as the influence of one variable, or stock, on another, so that the affected variable is caused to deviate from uniform change (Hayward, 2015). Such social forces include self forces. In the limited enthusiasm model, figure \ref{fig6.fig}, social forces are identified through the connections from stocks to the flows of the affected stocks. Thus there are four forces on $U$, two associated with loops \small\textsf{R1} \normalsize and \small \textsf{B2}, \normalsize  one from $U$ associated with an unnamed loop via $\Lambda_B$ and one from $A$ not associated with any loop; and three forces on $A$, associated with loops \small\textsf{R1}, \normalsize  \small \textsf{B2}  \normalsize and \small \textsf{B3}. \normalsize The forces on $B$ come from $U$, via $\lambda_B$;   a connection from $A$, also  via $\lambda_B$; and another connection from $A$ via $L_e$. None of these forces are part of a loop as there is no feedback from $B$ in the model, though the force via $L_e$ is the equal and opposite reaction to \small \textsf{B3} \normalsize on $A$ (Hayward \& Roach, 2018).

The social forces can be computed analytically by placing the differential equations in causally connected form using the procedure from Hayward and Roach (2018). From the model in figure \ref{fig6.fig} the causally connected differential equations of the limited enthusiasm model are:
\begin{eqnarray}
\frac{\dd U}{\dd t} &=& -\frac{gC_p}{\tau_a N}U_{\underline{p_UC_a\lambda_a\Lambda_A}}A_{\underline{\Lambda_A}}   -\frac{(1-g)C_p}{\tau_a N}U_{\underline{p_UC_a\lambda_a\Lambda_B}}A_{\underline{\Lambda_B}} \label{ccle.1} \\     
\frac{\dd A}{\dd t} &=& \frac{gC_p}{\tau_a N}U_{\underline{p_UC_a\lambda_a\Lambda_A}}A_{\underline{\Lambda_A}} -\frac{A_{\underline{L_e}}}{\tau_a}  \label{ccle.2} \\
\frac{\dd B}{\dd t} &=& \frac{(1-g)C_p}{\tau_a N}U_{\underline{p_UC_a\lambda_a\Lambda_B}}A_{\underline{\Lambda_B}} + \frac{A_{\underline{L_e}}}{\tau_a}\label{ccle.3}         
\end{eqnarray}
Each variable on the right hand side is annotated with the elements in each causal pathway as underlined subscripts. Analysis will be confined to the interpretation of the forces on the enthusiasts $A$ as it is their reproductive activity that drives the growth of the church. In general in a system dynamics model, although causal pathways are unique, the identification of feedback loops need not be, and there may be a number of independent loop sets. However in the limited enthusiasm model the three forces on $A$ are all identified uniquely with loops, thus (\ref{ccle.2}) can be rewritten with the loops names in the subscripts:
\begin{equation}
\frac{\dd A}{\dd t} = \frac{gC_p}{\tau_a N}U_{\underline{\mathrm{B2}}}A_{\underline{\mathrm{R1}}} -\frac{A_{\underline{\mathrm{B3}}}}{\tau_a} \label{ccle.2a}
\end{equation}

The impacts of the three forces are computed from (\ref{ccle.2a}) using pathway differentiation (Hayward \& Roach, 2018):
\begin{eqnarray}
\St_{\underline{\mathrm{R1}A}} &=&\left. \frac{\partial \dot{A} }{\partial A} \right\|_{\underline{\mathrm{R1}}}  = \frac{gC_p\bar{U}}{\tau_a}\label{impact.1}\\
\St_{\underline{\mathrm{B2}A}} &=&\left. \frac{\partial \dot{A} }{\partial U} \right\|_{\underline{\mathrm{B2}}} \times \frac{\dot{U}}{\dot{A}} = \frac{gC_p^2\bar{A}\bar{U}}{\tau_a(1-R_p\bar{U})}\label{impact.2} \\
\St_{\underline{\mathrm{B3}A}} &=&\left. \frac{\partial \dot{A} }{\partial A} \right\|_{\underline{\mathrm{B3}}}  = -\frac{1}{\tau_a}  \label{impact.3}
\end{eqnarray}

The impact of a force on a stock is the ratio of the acceleration of that stock (due to the force) with the rate of change. The sign of the impact matches the polarity of the forces effect on the variable, positive for reinforcing and negative for balancing. Thus the impact of the reinforcing loop \small $\mathrm{R1}$,  \normalsize is always reinforcing $\St_{\underline{\mathrm{R1}A}}>0$, and that of loop \small $\mathrm{B3}$  \normalsize is always balancing $\St_{\underline{\mathrm{B3}A}}<0$. This result follows from both loops being first order on $A$ (Hayward \& Boswell, 2014). However, loop \small $\mathrm{B2}$  \normalsize changes the polarity of its impact on $A$, balancing if $\bar{U}$ is above the threshold $\bar{U}_{\mathrm{revival}} =  1/R_p$ (using (\ref{rev.2})) $\St_{\underline{\mathrm{B2}A}}<0$, and reinforcing, $\St_{\underline{\mathrm{B2}A}}>0$,  when below the threshold. This loop, \small $\mathrm{B2}$,  \normalsize is able to change polarity on the stock $A$ as it is effectively second order, with the change being matched by  a change of the loop's polarity on $U$, through which it also passes (Hayward \& Boswell, 2014). The negative gain of the loop is thus preserved. The loop \small $\mathrm{B2}$,  \normalsize the resistive force of the diminishing unbelieving population on the conversion efforts of the enthusiasts, is key to why the generation of enthusiasts eventually falls and that growth in the church eventually stops. 

\begin{figure}[!ht]
          \begin{center}
   \includegraphics[height=7.0cm] {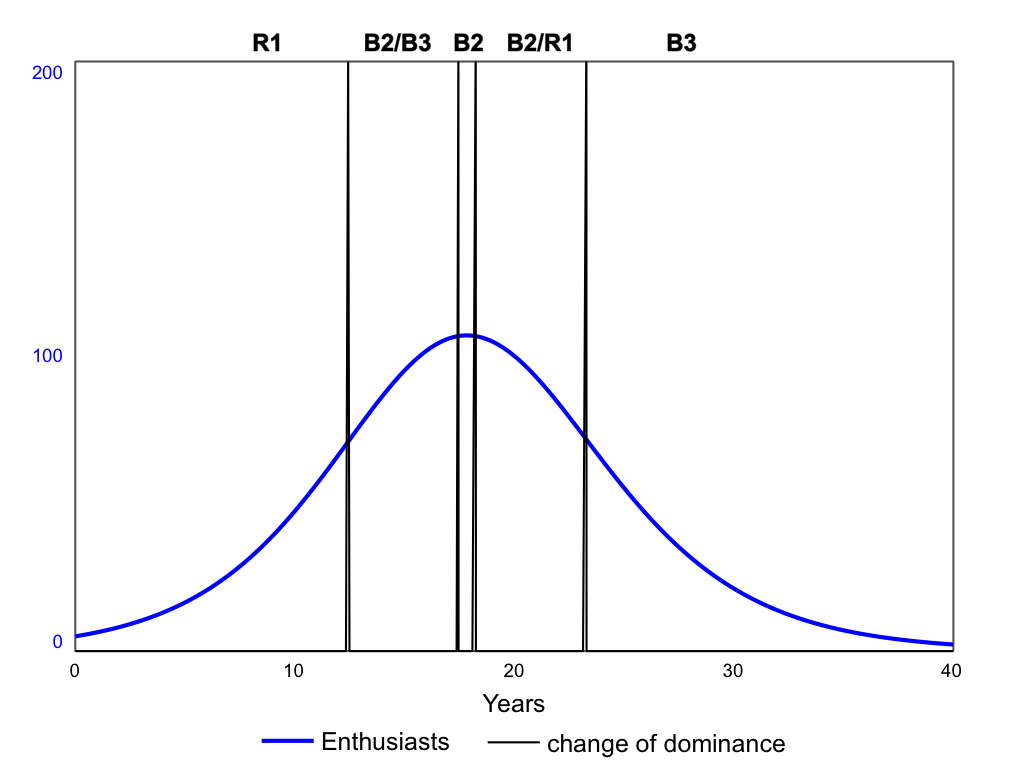}
       \end{center}
    \vspace{-20 pt}
    \caption{\small{Loop impact dominance (social forces) on enthusiasts, $A$. $N=50,000$, $C_0=A_0+B_0= 100$, $A_0/C_0 = 0.05\%$, $C_p=2.2$, $\tau_a = 0.4$ years, and $g=0.5$.} } \label{fig12.fig}
 \end{figure}

\begin{figure}[!ht]
          \begin{center}
   \includegraphics[height=7.0cm] {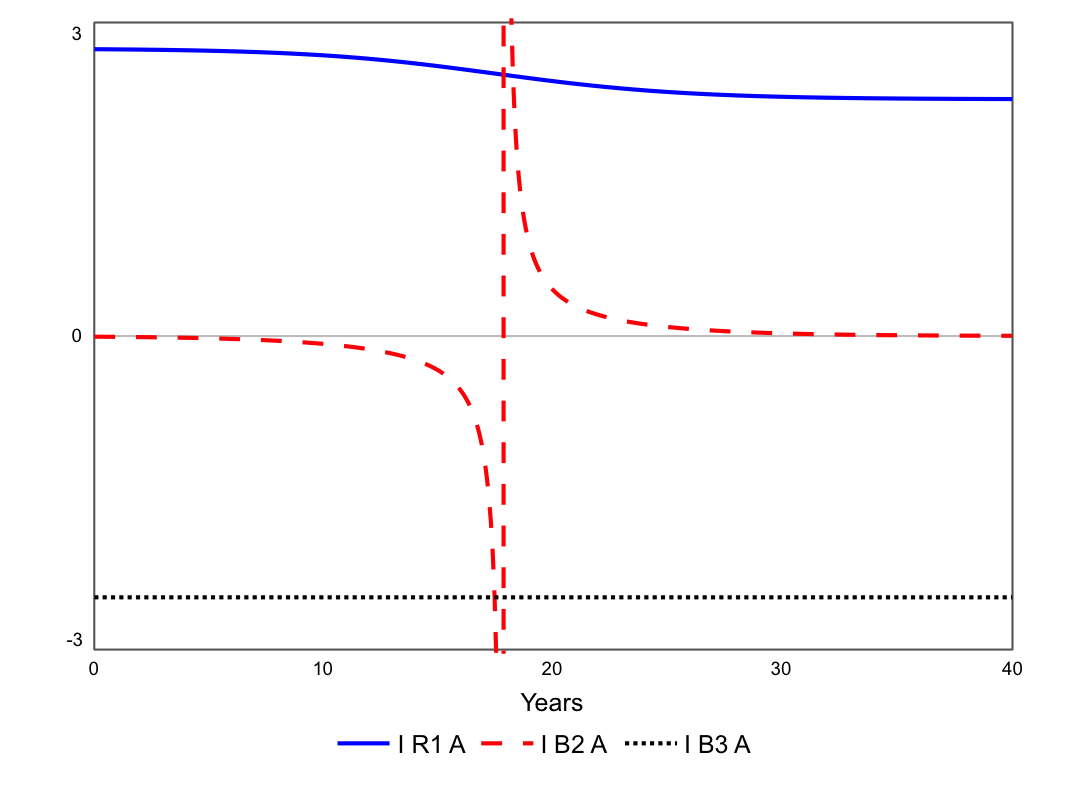}
       \end{center}
    \vspace{-22 pt}
   \caption{\small{Loop impact dominance (social forces) on enthusiasts, $A$. $N=50,000$, $C_0=A_0+B_0= 100$, $A_0/C_0 = 0.05\%$, $C_p=2.2$, $\tau_a = 0.4$ years, and $g=0.5$.} }  \label{fig13.fig}
 \end{figure}

Figure \ref{fig12.fig} marks the transitions between periods of loop dominance on the enthusiasts $A$. The regions indicate which loop, i.e. social force, of combination of loops, is responsible for curvature in the graph of $A$ against time. The accelerating phase of the revival growth in enthusiasts is dominated by the reinforcing loop  \small $\mathrm{R1}$.  \normalsize Thus, the activity of enthusiasts creating more enthusiasts drives the growth of the church. This phase lasts until 12.5 years when the combination of  \small $\mathrm{B2}$  \normalsize and  \small $\mathrm{B3}$  \normalsize exceeds the impact of \small $\mathrm{R1}$.  \normalsize Thus, the revival growth is opposed by the combination enthusiasts losing their enthusiasm,  \small $\mathrm{B3}$,  \normalsize and the resistance of the diminishing unbelieving pool making conversions harder to achieve,  \small $\mathrm{B2}$.  \normalsize Figure \ref{fig13.fig} shows the impacts of the three loops over time. Although the impact of \small $\mathrm{B3}$  \normalsize at 12.5 years is still numerically less than that of \small $\mathrm{R1}$,  \normalsize the increasing numerical impact of \small $\mathrm{B2}$  \normalsize is sufficient to enable \small $\mathrm{R1}$,  \normalsize to be counteracted. With $\St_{\underline{\mathrm{B2}A}} \propto \bar{A}/ (\bar{U}^{-1} - R_p)$, the resistance of \small $\mathrm{B2}$  \normalsize continues to increase numerically as $\bar{U} \rightarrow 1/R_p$ from above. With $\St_{\underline{\mathrm{R1}A}} \propto \bar{U}$, the impact of the enthusiasts drops as the unbelieving pool drops, and thus they cannot recover their initial accelerating revival growth. 

Due to the singularity of the impact $\St_{\underline{\mathrm{B2}A}}$ at $\bar{U} = 1/R_p$, then \small $\mathrm{B2}$  \normalsize alone causes the enthusiasts to move from growth to decline, phase \small $\mathrm{B2}$  \normalsize in figure \ref{fig12.fig}. $\St_{\underline{\mathrm{B2}A}}$ now has positive polarity, figure \ref{fig13.fig}, and thus combines with loop \small $\mathrm{R1}$  \normalsize to accelerate the decline of the enthusiasts. Eventually both loops decline, figure \ref{fig13.fig}, until the impact of \small $\mathrm{B3}$  \normalsize dominates, slowing the decline in $A$.

The key to  revival succeeding is minimising the resistance from unbelieving community by having contact with as large a community as possible, that is increasing $\bar{U}$. $\St_{\underline{\mathrm{B2}A}}$ remains smaller for longer, and $\St_{\underline{\mathrm{R1}A}}$ declines slower. It is suspected that many   churches that fail to grow, and  decline, do so because they have contact with too small a sub-population. That is $U+A+B < N$ due to a category of unbelievers with whom the enthusiasts have no effective contact. They may have many contacts but it is not sufficient  for epidemic spreading, thus even if a few enthusiasts become revived, they are unable to reproduce fast enough  and thus revival in the community  eludes them. The Toronto Blessing of 1994 onwards is an example of a revival movement within the church that had no direct observable effect on the unbelieving population (Poloma, 2003). However a subsidiary of that movement, the Alpha Course, has succeeded in having revival impact on the wider community with over 3 million people in the UK having participated in the course, and many going on to church membership.

\section{Conclusion}

\subsection{\label{Conclusion}Main Conclusions}

The primary aim of this paper was to investigate whether population models,
in particular the epidemic model with its spread by contact and limited
infectious period, could be used to model a growing church. As shown in
sections \ref{The Simple Church Growth} and \ref{Numerical Solutions} the
results of the model do exhibit typical church growth behaviour,
particularly that seen during revival. Further, the construction of the
equations can be explained in terms of the dynamical processes that take
place between unbelievers and the two categories of believers, albeit a
highly simplified model. The mass action principle (personal contacts) is
well suited to modeling the dynamics of conversion, and provides the
typically S-shaped behaviour found in the growth of churches. The limited
duration for the enthusiastic, or recruitment, phase effectively prevents
the church from growing to the whole population. In general it can be
concluded that the epidemic model is a suitable starting point for
investigating the dynamics of church growth.

A number of specific conclusions can also be drawn from investigating the
effects of changing parameters and initial conditions:

\begin{enumerate}
\item  Improving the effectiveness of believers in evangelism has a more
significant effect than increasing the number of evangelisers. Whether this
has any implications for evangelism training is not clear. It may not be
very easy to improve a person's evangelistic effectiveness. However it does
help explain why revivals can start with such low numbers of infected
believers. If the effective conversion rate increases by only a modest amount,
either by changes in the enthusiasts, or changes in the
unbelievers' receptivity, then growth can very quickly take off.

\item  If the Christian church is a small proportion of the whole
population, and enthusiasts a small proportion of the church, then
revival growth is possible but its build up tends to be slow. The bulk of
the converts in a revival come in its middle period, given that all
parameters remain the same. This is a direct result of the logistic
behaviour of the growth. Thus if a revival lasts $20$ years the number of
converts over the first $5$ years may not be that noticeable. The 
phenomena referred to as the ``Toronto Blessing'' had many of the hallmarks
of a revival, but in its early years no large number of conversions had been
unambiguously measured in the wider church. As noted earlier this has led
many to refrain from calling it a revival. However, given the low numbers in
the church in western countries, any revival is likely to be long term, with
the characteristic slow build up, as such it is too early see substantial
growth in a religious phenomena that is less than five years old. In the case of the Toronto blessing, if other movements it has influenced, such as the Alpha Course, are included then in the 24 years since the movement started,   many  conversions have eventually accumulated.

\item  When the church is a larger proportion of the population, a higher
conversion rate among enthusiasts is needed for revival growth to
occur. Given other parameters remain the same such revivals are shorter than
those in countries where the church is weak. The larger the revival the
faster it occurs, thus for a large revival the number of conversions at its
peak can be very dramatic.

\item  Revivals can burn out for dynamical reasons, i.e. the number of
susceptibles falls to a level where the conversion rate proves inadequate to
sustain a revival. The longer revival growth continues, the harder it is to
keep going because there are less unbelievers. This is a direct result of
the spread by contact, with the enthusiasts effort being increasingly
``wasted'' on those already converted.

\item[ ]  However the growth will stop before all unbelievers are reached.
It does not gradually get slower until everyone is converted, this being a
direct result of limiting the enthusiastic, or recruitment, phase of a
believer. Of course revivals can end for other reasons, such as infected
believers being ``less infected''. In this model this is represented by
changes in the parameters of the model.

\item  Revivals do not end because the people involved become more
secularised. Typical timescales for enthusiastic periods are very short,
whereas secularisation is a long term process. It is quite possible the lack
of enthusiasm and the inability to win converts contributes to the
secularisation of the church (Stark and Bainbridge, 1985, p.364). 

\item  Because the model concentrates on the dynamics of the growth, it is
independent of whether the underlying factors which effect growth are
institutional, social or theological. Of course if any of parameters change
in time then the relative merits of these factors would need to be
investigated.
\end{enumerate}

\subsection{Further Work}

The limited enthusiast church model needs to be extended to account for other mechanisms
for growth, as outlined in section \ref{Use of Basic}.

\begin{enumerate}
\item  Growth through births and deaths are needed to model long term
behaviour. As well as the supply of unbelievers being replenished, the
children of believers may automatically enter the church without ever being
classed as unbelievers or infected. Children of believers often have less
interest in enthusiastic religion than their parents, are more socialised,
and hence push the church towards ineffectiveness (Stark and Bainbridge, 1985, pp.24, 157--165)\endnote{These features were added in Hayward, 2005.}.

\item  Transfer growth exists between individual congregations, and between
denominations. This opens up the possible need to model competition between
different types of churches. This was characteristic of the early spread of
Christianity as it competed with paganism  (Stark, 1996, p.191), and is
particularly true in the USA and Europe which are highly pluralistic
religious markets (Fink and Stark, 1992; Iannaccone, 1991; Stark and Bainbridge, 1985, p.353; Stark and Iannaccone, 1994).

\item  Churches may decline through reversion, i.e. some believers revert
back to being unbelievers\endnote{This feature was added in Hayward, 2005.}.

\item  Inactive believers can become enthusiastic again, i.e. they are
re-infected from existing enthusiasts. This process is essential if the
early phase of a revival is to be modelled\endnote{This feature was added in Hayward, 2010.}.

\item  Further categories of people can be considered. There may be more
than one category of unbeliever some less resistant than others to
conversion, thus creating possible pools of potential recruits (Stark and Bainbridge, 1985, p.351f). Secondary conversions such as the spouses of initial
converts may have far less enthusiasm  (Stark, 1996, p.100) and be a
separate category of believer.

\item  The susceptible pool is not strictly homogeneous as converts often
come through existing social networks (Stark and Bainbridge, 1985, p.312f).
\end{enumerate}

The model will be extended in future publications.

\section*{Acknowledgements}

The author would like to thank anonymous referees and the editor for many
helpful remarks concerning the original, 1999,  version  of work, (Hayward, 1999).

\begingroup
\def\enotesize{\small}
\makeatletter
\def\enoteformat{\rightskip\z@ \leftskip1em \parindent=0em
\leavevmode\llap{\hbox{\@theenmark.~}}}
\makeatother
\theendnotes
\endgroup

\section*{\emph{References}}
\begin{description}
\item[] Anderson R.M. (1988), The Epidemiology of HIV Infection, \emph{Journal of the  Royal Statistical Society A}, \textbf{151 part1}.

\item[] Anderson R.M. and May R.M. (1987), \emph{Infectious Diseases in Humans: Dynamics and Control}, OUP.
 
\item[] Bailey N.T.J. (1975), \emph{The Mathematical Theory of Infectious Diseases and its Applications}, Griffen, London.

 \item[] Banks R.B. (1994), \emph{Growth and Diffusion Phenomena}, Springer-Verlag.

\item[] Bartholomew D.J. (1967), \emph{Stochastic Models for
Social Processes}, Wiley, New York.

 \item[] Bartleman F. (1980), \emph{Azusa Street: The Roots of Modern Day Pentecost}, Logos International.

 \item[] Berger P. (1969), \emph{The Sacred Canopy: Elements of a
Sociological Theory of Religion}, Anchor NY.

\item[]Berger P. (1970), \emph{A Rumour of Angels: Modern Society
and the Re-discovery of the Supernatural}, Anchor NY.

\item[]Braun M. (1975), \emph{Differential Equations and their
Applications}, (Springer Verlag).

\item[]Brierley P. (1991), \emph{Christian England -- What the
English Church Census Reveals}, MARC Europe, London.

\item[]Brierley P. (1993), More Down Than Up, \emph{Quadrant}, Nov.
1993, Christian Research Organisation.

\item[]Bulan S. and Bulan-Dorai L. (2004), \emph{The Bario Revival},
Home Matters Network, Kuala Lumpur, Malaysia.

\item[]Burden D.L. and Faires J.D. (1988), \emph{Numerical Analysis},
PWS-Kent, Boston.

\item[]Butler B. (1976), \emph{Hill Ablaze},
Hodder and Stoughton.

\item[]Coleman J.S. (1964), \emph{Introduction to Mathematical
Sociology}, The Free Press of Glencoe NY.

\item[]Coleman R. and Litfin A.D. (1995), \emph{Accounts of a Campus Revival -- Wheaton College 1995}, Harold Shaw Publishers.

\item[]Cross W.R. (2015), \emph{ The Burned-Over District: The Social and Intellectual History of Enthusiastic Religion in Western New York,1800-1850}, Cornell University Press.

\item[]Davies E. (2004), \emph{The Beddgelert Revival}, Bryntirion Press, Wales, UK.

\item[]DeLoriea R. (1997), \emph{Portal in Pensacola}, Revival Press, Destiny Image, PA.

\item[]Doyle R.T. and Kelley S.M. (1979), Comparison of
Trends in Ten Different Denominations, pp. 144--159, in ``\emph{Understanding
Church Growth and Decline 1950--1978}'' edited by Hoge D.R and Roozen D.A.,
Pilgrim Press.

\item[]Edwards B.H. (1990), \emph{Revival}, Evangelical Press.

\item[]Edwards J. (1984), \emph{Jonathan Edwards on Revival}, Banner of Truth Trust.

\item[]Evans E. (1967), \emph{Revival Comes to Wales}, Evangelical Press of Wales.

\item[]Evans E. (1969), \emph{The Welsh Revival of 1904},
Evangelical Press of Wales.

\item[]Evans E. (1985), \emph{Daniel Rowland and the Great Evangelical Awakening}, Banner of Truth.

 \item[]Fink R. and Stark R.(1992), \emph{The Churching of America
1776 -- 1990: Winners and Losers in our Religious Economy}, Rutgers University
Press.

\item[] Forrester J.W. (1961), \emph{Industrial Dynamics}, Pegasus Communications, Waltham MA.

\item[]Gaustad  E.S. (1968), \emph{The Great Awakening in New England}, Quadrangle Books. 

\item[]Granoveter M. and Soong R. (1983), Threshold
Models of Diffusion and Collective Behavior, \emph{Journal of  Mathematical  Sociology}, \textbf{9},165--179.

\item[] Green M. (1990), \emph{Evangelism Through the Local
Church}, Hodder and Stoughton, London.

\item[]Griffin S.C. (1992), \emph{A Forgotten Revival}, Day One Publications. 

\item[]Hadaway C.K. (1993a), Do Church Growth
Consultations Really Work? in ``\emph{Church and Denominational Growth}'', edited
by Roozen D.A. and Hadaway C.K., (1993), 149--154.

\item[]Hadaway C.K. (1993b), Is Evangelistic Activity
Related to Church Growth? in ``\emph{Church and Denominational Growth}'', edited by
Roozen D.A. and Hadaway C.K., (1993), 169--187.

\item[] Hamer W. H. (1906), Epidemic Disease in England,
\emph{The Lancet}, \textbf{i}, 733--9.

\item[]Hattaway P. (2006), \emph{From Head-Hunters to Church Planters: An Amazing Spiritual Awakening in Nagaland},  IVP.

\item[]Hayward J. (1999), Mathematical Modeling of Church Growth, \textit{Journal of Mathematical Sociology}, \textbf{23(4)}, pp.255--292. 

\item[] Hayward J. (2000), Growth and Decline of Religious and Subcultural Groups, in \emph{Proceedings of the 18th International Conference of the Systems Dynamics Society}, Bergen, Norway.

\item[] Hayward J. (2002), A Dynamical Model of Church Growth and its Application to Contemporary Revivals, \emph{Review of Religious Research}, \textbf{43(3)}, 218-241.

\item[]Hayward J. (2005), A General Model of Church Growth and Decline, \emph{Journal of Mathematical Sociology}, \textbf{29(3)}, pp.177--207.

\item[] Hayward J. (2010), Church Growth via Enthusiasts and Renewal, in \emph{Proceedings of the 28th International Conference of the Systems Dynamics Society}, Seoul, South Korea.

\item[]Hayward J. and Boswell G.P. (2014), Model Behaviour and the Concept of Loop Impact: A Practical Method. \emph{System Dynamics Review}, \textbf{30(1-2)}, pp.29--57.

\item[]Hayward, J. (2015). Newton's Laws of System Dynamics, in \emph{Proceedings of the 33rd International Conference of the Systems Dynamics Society}, Cambridge,  MA.

\item[]Hayward J. and Roach P.A.. (2018), Newton's Laws as an Interpretive Framework in System Dynamics, \emph{System Dynamics Review}, to be published, DOI 10.1002/sdr.1586.

\item[]Hethcote H.W.
(1994), A thousand and one epidemic models. In \textit{Frontiers
in Mathematical Biology}, ed. S.A. Levin. Berlin: Springer Verlag.

\item[]Hoge D.R. (1979), A Test of Theories of Denominational
Growth or Decline, pp.179--197, in ``\emph{Understanding Church Growth and Decline
1950--1978}'', edited by Hoge D.R. and Roozen D.A., Pilgrim Press.

\item[]Hoge D.R. and Roozen D.A.(Eds.) (1979), \emph{Understanding
Church Growth and Decline 1950--1978}, Pilgrim Press.

\item[]Hunt S. (2009),  \emph{History of the Charismatic Movement in Britain and the United States of America: The Pentecostal Transformation of Christianity}, Edwin Mellen Press Ltd.

\item[]Kelley D. (1986), \emph{Why Conservative Churches are Growing:
A Study in the Sociology of Religion with a New Preface for the ROSE
Edition}, Mercer University Press.

\item[]Iannaccone L.R. (1991), The Consequences
of Religious Market Structure, \emph{Rationality and Society}, \textbf{3}, (April), 156--177.

\item[]Iannaccone L.R. (1992) Religious Markets and
the Economics of Religion, \emph{Social Compass}, \textbf{39 (1)}, 123--131.

\item[]Iannaccone L.R. (1994), Why Strict Churches are
Strong, \emph{American Journal of Sociology}, \textbf{99(5)}, 1180--1211.

\item[]Iannaccone L.R., Olson P. and Stark R.
(1995), Religious Resources and Church Growth, \emph{Social Forces},  \textbf{74(2)},
705--731.

\item[]Inskeep K.W. (1993), A Short History of Church
Growth Research, in ``\emph{Church and Denominational Growth}'', edited by Roozen
D.A. and Hadaway C.K., (1993), 135--148.

 \item[]Kermack W.O. and McKendrick A.G. (1927), A
Contribution to the Mathematical Theory of Epidemics, \emph{Proceedings of the  Royal Society}, \textbf{A115},
700--21.

\item[] Kumar V. and Kumar U. (1992), Innovation
Diffusion: Some New Technological Substitution Models, \emph{Journal of Mathematical Sociology},
\textbf{17(2--3)}, 175--194.

\item[]Lloyd-Jones D.M. (1984), \emph{Joy Unspeakable}, Kingsway
Publications.

\item[]Lloyd-Jones D.M. (1986), \emph{Revival}, Marshall
Pickering.

\item[]Mahajan V., Muller E. and Bass F.M. (1990), New
Product Diffusion Models in Marketing: A Review and Directions for Research,
\emph{Journal of Marketing}, \textbf{54}, 1--26.

\item[]Martin D. (2001), \emph{Pentecostalism: The World Their Parish},  Blackwell: Oxford.

\item[]May R.M. and Anderson R.M. (1985), Endemic Infections
in Growing Populations, \emph{Mathematical Biosciences}, \textbf{77}, 141--156.

\item[]May R.M. and Anderson R.M. (1987), Transmission
Dynamics of HIV Infection, \emph{Nature}, \textbf{326}, 137--142.

\item[]McCallum H. Barlow N. and Home J. (2001), How
should pathogen transmission be modelled? \textit{Trends in
Ecology and Evolution}, \textbf{16:6}: 295--300.

\item[]McGavran D. (1963), \emph{Do Churches Grow?} World
Dominion Press, Reprinted by the British Church Growth Association 1991.

\item[]Miller A.S. and Nakamura T. (1996) On the
Stability of Church Attendance Patterns During a Time of Demographic Change:
1965--1988, \emph{Journal for the Scientific Study of Religion}, \textbf{35(3)}, 275--284.

\item[]Murray J.D. (1989), \emph{Mathematical Biology},
Springer-Verlag.

\item[]Neighbour R. (1990), \emph{Where Do We Go From Here?},
Touch Publications -- Houston.

\item[]The Open University (1988), \emph{Mathematics: A Third
Level Course}, M343 Applications of Probability, Unit 10 -- Epidemics.

\item[]Olson D.V.A. (1989), Church Friendships: Boon or
Barrier to Church Growth?, \emph{Journal for the Scientific Study of Religion},
\textbf{28(4)}, 432--447.

\item[]Orr J.E. (2000), The Outpouring of the Spirit in Revival and Awakening and its Issue in Church Growth, British Church Growth Association, republished by Church Growth Modelling. 

\item[]Peckham C. and Peckham M. (2004), \emph{Sounds From Heaven: The Revival on the Isle of Lewis 1949--1952}, Christian Focus Publications.

\item[]Pointer R. (1987), \emph{The Growth Book}, MARC Europe --
British Church Growth Association.

\item[]Poloma M.M. (2003),  \emph{Main Street Mystics: The Toronto Blessing and Reviving Pentecostalism}, Altmira Press. 

\item[]Ragget G.F. (1982), Modeling the Eyam Plague,
\emph{Bulletin of the Institute of Mathematics and its Applications}, \textbf{18}, 
221--226.

\item[]Riss R.M. (1988), \emph{A Survey of $20^{th}$ Century Revival Movements in North America}, Hendrickson Publishers.

\item[]Riss R.M. and Riss K. (1997), \emph{Images of Revival}, Destiny Image.

\item[]Robinson M. (1993), \emph{A World Apart}, Monarch/CPAS.

\item[]Robinson M. (1994), What Next After Toronto?,
\emph{Church Growth Digest}, \textbf{16(2)}, p15, British Church Growth Association.

\item[]Rogers E.M. (1995), \emph{Diffusion of Innovations},
(4th Ed.), The Free Press, New York.

\item[]Roozen D.A. and Hadaway C.K. (eds.) (1993), \emph{Church
and Denominational Growth: What Does (and Does Not) Cause Growth and
Decline}, Abingdon Press.

\item[]Ryle J.C. (1978), \emph{Christian Leaders of the 18th Century}, Banner of Truth.

\item[]Sharif M.N. and Ramanathan K. (1982), Polynomial
Innovation Diffusion Models, \emph{Technological Forcasting and Social Change}, \textbf{21},
301--323.

\item[]Stark R. (1996), \emph{The Rise of Christianity}, Princeton
University Press.

 \item[]Stark R. and Bainbridge W.S. (1985), \emph{The Future of
Religion}, University of California Press.

\item[]Stark R and Bainbridge W.S. (1987), \emph{Theory of
Religion}, Rutgers University Press.

\item[]Stark R. and Iannaccone L.R. (1994), A Supply Side
Re-Interpretation of the ``Secularisation'' of Europe, \emph{Journal for the
Scientific Study of Religion}, \textbf{33(3)}, 230--252.

\item[]Sterman J.D.
(2000), \textit{Business Dynamics: Systems Thinking and Modeling
for a Complex World}, Irwin/McGraw:Hill: New York.

\item[]Taylor S.
(2003), \textit{The Skye Revivals}, New Wine Press, UK.

\item[]Wagner C.P. (1987), \emph{Strategies for Church Growth},
MARC Europe -- British Church Growth Association.

\item[]Wallace A.F.C. (1966), \emph{Religion: An
Anthropological View}, New York Random House.

\item[]Ward K. and  Wild-Wood E. (2010), \emph{The East African Revival: Histories and Legacies}, Fountain Publishers, Kampala, Uganda.

 \item[]Warner R.S. (1993), Work in Progress toward a New
Paradigm for the Sociological Study of Religion in the United States,
\emph{American Journal of Sociology}, \textbf{98 (5)}, 1044--93.

\item[]Webber M.J. (1972), \emph{Impact of Uncertainty on
Location}, MIT Press.

\item[]Weisberger B.A. (1966) \emph{They Gathered at the River: The Story of the Great Revivalists and their Impact Upon Religion in America}, Quadrangle Books.

\item[]Williams J. (1985), \emph{Digest of Welsh Historical
Statistics}, Government Statistical Service HMSO.

 \item[]Wimber J. (1994), Season of New Beginnings,
\emph{Equipping the Saints}, \textbf{Fall 1994}, VMI International.
\end{description}

\clearpage
\appendix

\section{Glossary}
\subsection*{Dynamical variables (stocks and their densities):}
\begin{tabular}{ll}
$t$ & Time \\
$S$ & Number of susceptibles.  \\
$I$ & Number of infectives.  \\
$R$ & Number removed from epidemic.  \\
$N$ & Total number of population under consideration. \\
$U$ & Number of unbelievers.  \\
$A$ & Number of enthusiasts (active believers, infected believers). \\
$B$ & Number of inactive believers. \\
$\bar{U}$ & Proportion of unbelievers in total population. \\
$\bar{A}$ & Proportion of enthusiasts in total population. \\
$\bar{U}_{0}$ & Initial proportion of unbelievers in total
population. \\
$\bar{A}_{0}$ & Initial proportion of enthusiasts in total
population.\\
\end{tabular}

\subsection* {Rates of change (flows):}
\begin{tabular}{ll}
$i$ & Infection rate. \\
$c$ & Cure rate.  \\
$\Lambda_A$ & Rate of conversion of unbelievers to enthusiasts.  \\
$\Lambda_B$ & Rate of conversion of unbelievers to inactive believers.  \\
$L_e$ & Rate of transition of enthusiasts to inactive believers as they lose enthusiasm.  \\
\end{tabular}

\subsection* {Parameters and auxiliaries (converters):}
\begin{tabular}{ll}
$\beta $ & Contact rate. \\
$\tau$ & Duration of infection. Infectious period.\\
$\gamma $ & $1/\tau$. \\
$\lambda_i$ & Per capita rate of infection. \\
$n_S$ & The number of susceptibles who are infected through contact with an infective
\\
&   during their whole infectious period. \\
$n_N$ & The number of susceptibles who are infected through contact with an infective \\
&   during their whole infectious period  given the whole population is susceptible. \\ 
$p_S$ & The probability  one infective contacts a susceptible in a homogeneous population.\\
$\tau_a$ & Duration of enthusiastic phase. \\
$\lambda_a$ & Per capita rate of conversion. \\
$C_a$ & The number of conversions through one enthusiast during their whole enthusiastic 
\\
&period.\\
$C_p$ & The number of conversions through one enthusiast during their whole enthusiastic 
\\ &period given the whole population are unbelievers. The conversion potential.\\
$R_p$ & The number of enthusiasts made through one enthusiast during their whole  
\\ &enthusiastic period given the whole population are unbelievers. The reproduction\\ & potential.\\
$p_U$ & The probability  one enthusiast contacts an unbeliever in a homogeneous population.\\
g & The fraction of new converts who become enthusiasts.\\
$g^{\prime} $&  $1-g$.  The fraction of new converts who become inactive believers.\\
\end{tabular}

\section{Terminology}

\begin{itemize}
\item[ ]  \textbf{Believer: }A member of a church, also referred to as a
\textbf{church member} and a \textbf{Christian} in the paper. How membership
is defined, or whether all members really are believers, or whether all
attenders are members, would require more sophisticated models. A believer
is also called a \textbf{convert} if they have recently become a believer.

\item[ ]  \textbf{Unbeliever: }A person who is not a member of a church,
\textbf{a non-Christian}.

\item[ ]  \textbf{Enthusiast: }A believer with a much higher activity
in passing on the faith to unbelievers. Also called \textbf{active
believer}, or \textbf{infected believer}. They may be engaged in a systematic
program of evangelism i.e. an \textbf{evangeliser}, or they may be caught up
in a ``revival-type'' behaviour causing them to be an effective witness. It
is assumed in this paper that they are new-converts.

\item[ ]  \textbf{Inactive believer: }Those removed from the process of
winning new converts, at least by comparison with infected believers

\item[ ]  \textbf{Conversion: }The process by which an unbeliever becomes a
believer. This is represented as a transfer from category $U$ to
category $A$. All new converts are assumed infected immediately.
Also referred to in this report as being as an unbeliever being brought or
led to Christ.

\item[ ]  \textbf{Church: }This may refer to an individual \textbf{congregation}, i.e. a local group of Christians who meet together on a
Sunday under a common leadership. It may also refer to all the people who
belong to Christian congregations throughout a country, or the world,
regardless of denomination -- i.e. the \textbf{whole church}. The size of the church can be measured by attendance at the main services or by an official system of membership.
\end{itemize}

\section{Revised Notation}
Hayward (1999),on which this paper is based, was the first publication on the mathematics of church growth, based on a technical report of 1995. Later publications have undergone a change of notation, and a greater use of the fixed contacts model, partly as a result of using the System Dynamics methodology. These notational changes are included in this 2018 paper.

\end{document}